\documentclass{article}
\usepackage{cite}

\usepackage[english]{babel}

\usepackage[table]{xcolor}
\usepackage{tabularx,booktabs}
\usepackage{colortbl}
\usepackage{amsmath}
\usepackage{graphicx}
\usepackage{subcaption}
\usepackage[colorlinks=true, allcolors=blue]{hyperref}
\usepackage{longtable}
\usepackage{pgfplots}
\usepackage{pgfplotstable}
\usepackage{tikz}
\usepackage{authblk}
\usepackage[a4paper, margin=1.2in]{geometry}
\usepackage{color}
\usepackage{array}
\usepackage{xcolor}
\usepackage{pifont}
\usepackage{multirow}

\newcommand{\xmark}{\ding{55}}%
\definecolor{codegreen}{rgb}{0,0.6,0}
\definecolor{codegray}{rgb}{0.5,0.5,0.5}
\definecolor{codepurple}{rgb}{0.58,0,0.82}
\definecolor{backcolour}{rgb}{0.95,0.95,0.92}
\usepackage{listings}
\usepackage{listings-golang}
\lstdefinestyle{mystyle}{
backgroundcolor=\color{backcolour},
commentstyle=\color{codegreen},
keywordstyle=\color{magenta},
numberstyle=\tiny\color{codegray},
stringstyle=\color{codepurple},
basicstyle=\ttfamily\footnotesize,
breakatwhitespace=false,
breaklines=true,
captionpos=b,
keepspaces=true,
numbers=left,
numbersep=5pt,
showspaces=false,
showstringspaces=false,
showtabs=false,
tabsize=2
}

\usepackage{booktabs}
\usepackage{rotating}
\usepackage{multicol}
\usepackage{subcaption}
\usepackage{caption}
\usepackage{amsmath}
\usepackage{graphicx}
\PassOptionsToPackage{hyphens}{url}
\usepackage{url}

\usepackage{csvsimple}
\usepackage{amssymb}
\PassOptionsToPackage{inline}{enumitem}
\usepackage{enumitem}

\usepackage{pgfplots}

\pgfplotsset{compat=newest}
\definecolor{Highlight}{rgb}{0.68,1,0.68}
\usepackage{pgf-pie}
\usepackage{changes}
\providecommand{\keywords}[1]{\textbf{\textit{Keywords:}} #1}

\begin{document}
\title{Exploring the connection between coding habits and cognitive styles in malware developers}

\author[1]{Vasilis Vouvoutsis}
\author[1,2]{Constantinos Patsakis}
\author[3]{Fran Casino}

\affil[1]{Department of Informatics, University of Piraeus, Piraeus, Greece}
\affil[2]{Athena Research Centre, Greece}
\affil[3]{Dept. of Computer Engineering and Mathematics, Universitat Rovira i Virgili, Catalonia, Spain}
\date{}
\maketitle

\begin{abstract}
Malware research primarily studies the results, the methods, and the impact. Even from an offensive security perspective, what is examined is the method, not the development strategy of the offender.
This study investigates the behavioral signatures and coding patterns embedded in the malware source code. By analyzing a large corpus of leaked malware code and comparing it with carefully selected benign open-source software, we apply static application security testing and compute multiple software metrics. Based on cognitive psychology and criminological theories, our work interprets differences in code structure and quality as behavioral indicators, reflecting distinct motivational structures, risk tolerances, and development strategies of malware authors compared to benign software developers. Our findings reveal that malware code is generally smaller, less documented, and exhibits higher cyclomatic complexity per function, with reduced use of abstraction mechanisms such as classes and closures. Vulnerability analysis further reveals that malware exhibits more issues of the types that benign code typically avoids, suggesting a minimal investment in secure development practices. These patterns imply a development style optimized for expedience, operational secrecy, and evasion rather than long-term maintainability. Nonetheless, the code quality metrics indicate that it does not deviate significantly from benign software enough to be distinctive. By framing code metrics as proxies for behavioral signals and strategic choices, we demonstrate how quantitative software analysis can enrich behavioral cybersecurity research, offering new insights into the practices and priorities of malware developers. Our results pave the way for further research in the behavioral profiling of cyber offenders.
\end{abstract}

\section{Introduction}
For decades, programmers have been writing code in various programming languages to perform multiple tasks with varying levels of complexity. Although such a task is highly technical and the writing is not performed in a human language, it does not lack a personal touch in any way. Cognitive style and personality shape different stages of the programming process, from problem framing to debugging, which can be mapped onto observable coding habits \cite{bishop1995cognitive}. Essentially, programmers use the code to express themselves, so it can be considered a form of speech \cite{cox2012speaking}. In fact, in the United States, source code has been argued and treated as protected speech, see the famous series of cases of Bernstein vs. the United States \cite{collins1997speaking}. Moreover, personal traits are also explored in stylometry in terms of code to enable the identification of malware authors \cite{krsul1997authorship,kalgutkar2019code,knochel2024text}.
Based on the above, one could ponder whether such traits can be identified in specific groups of people. In this line of thought, we want to examine the programming behavior of malware authors and investigate whether they exhibit significant differences from what we could consider 'traditional' programmers, that is, authors of benign code, that is, benign software intended not to cause harm. This analysis assesses, for instance, vulnerabilities, weaknesses, and other particularities in code and what they reveal about development practices. In this regard, we do not target attribution or assess individuals; rather, we determine how malicious code deviates from the source code of benign software.
As a result, this work is positioned at the intersection of malware analysis and software security analysis, since malware analysis traditionally focuses on understanding malicious behavior or detecting malware, while developers typically use static application security testing (SAST) to find bugs in benign applications.

While there is a plethora of malware repositories, they come in the form of compiled code, which is also packed and obfuscated. Therefore, even by reverse engineering the binaries, the end result would be far from the original. To overcome this limitation, we examine \emph{leaked} malicious code in public repositories of various malware and programming languages to identify differences from benign code and to examine whether malware authors have different mindsets and strategies compared to developers of benign software. The primary reason for this scoping is that malware remains one of the most pressing cybersecurity threats, and its increasing impact is driving the need for sophisticated detection and analysis techniques. Although static and dynamic behavior analyses have long been a cornerstone of malware research, primarily for classification, source code analysis offers invaluable insights by examining source code without execution \cite{bhutani2024analysing}. Thus, this work builds on the identification of structural, functional, and developmental characteristics of malicious software. Moreover, by using SAST, researchers can complement their findings to discover vulnerabilities, analyze coding patterns, and gain insight into malware's own vulnerabilities or poor coding practices that could potentially be exploited for defensive purposes, as in the case of Malvuln \cite{malvuln}.

Moreover, this work is situated at the intersection of software engineering and behavioral science, drawing on theories from cognitive psychology and criminology to interpret coding practices as behavioral artifacts. In particular, we consider that programming is an expression of cognitive style, problem-solving strategy, and developer priorities, as proposed by Bishop-Clark's framework on cognitive style and programming behavior \cite{bishop1995cognitive}. From a cognitive load perspective, malware authors operate under conditions of secrecy, time pressure, and, of course, legal risk, which may lead them to adopt coding behaviors that benign developers would not.

In addition, criminological theories of rational choice and opportunity structure \cite{cornish2014reasoning} suggest that offenders, including malware developers that we study in this work, make calculated decisions to maximize success while minimizing effort and detection risk. The structure of malware code, as revealed in our analysis, appears to reflect such priorities. Most malware appears simple, often consisting of quickly developed scripts with limited maintainability and documentation, which minimize the forensic footprint and maximize deployability under adversarial conditions. In this sense, the technical differences between malware and benign software are not merely engineering artifacts, but also behavioral traces and manifestations of different motivational structures, risk tolerances, and ethical orientations. Hence, by analyzing these differences quantitatively, we aim to provide behavioral insights into the development practices of malware authors as a distinct population of programmers.

In fact, this framing aligns with prior work on stylistic forensics and software authorship attribution, which recognizes that code can be used as a behavioral fingerprint, encoding individual and group-level differences in cognitive orientations \cite{kalgutkar2019code}. As a result, our study extends this perspective by focusing on malware as a domain where these behavioral signals may be particularly pronounced due to its illicit nature.

\subsection*{Contribution and Plan of the Article}
Despite the advances in state-of-the-art technology, to the best of our knowledge, no prior study has combined code metrics with SAST to study malicious code. In this regard, we propose a methodology that uses SAST tools and code metrics to measure various code attributes, including size and complexity, but also identify vulnerabilities and weaknesses. By examining both benign and malicious software code, SAST can reveal structural and functional differences, providing crucial context to understand how malware is constructed and behaves. Moreover, it provides insight into what developers prioritize when writing software for different purposes. The present work fills the existing gap by analyzing 658 leaked projects across seven programming languages, computing size, complexity, and maintainability indicators alongside full Common Weakness Enumeration (CWE) enumerations, and contrasting the results with 249 carefully matched open-source projects.

To better understand the mentality and priorities of malware developers, we establish the following research questions that we try to answer by performing thorough measurements on our dataset.
\begin{itemize}
\item Are there measurable structural differences between malware and benign source code, particularly in terms of code size, complexity, and maintainability?
\item Can patterns or clusters of CWEs be identified that differentiate malware code from benign software?
\item To what extent does malware source code deviate from industry-standard quality and maintainability metrics, such as the Maintainability Index (MI), Halstead metrics, and Cyclomatic Complexity?
\item How can these insights derived from static code analysis improve malware disruption efforts?
\item Can software engineering metrics effectively serve as behavioral artifacts to distinguish the underlying motivational structures and risk tolerances of malware authors from those of benign developers?
\end{itemize}
To answer the above, throughout our analysis, we interpret structural code metrics and identified vulnerabilities not only as engineering artifacts but also as quantitative proxies for the cognitive strategies and behavioral adaptations of developers operating under different incentive structures and goals that malware and benign code developers have.

The remainder of the article is structured as follows. In Section \ref{sec:back}, we provide background on the main concepts and metrics to better understand the methods discussed in the article. Section \ref{sec:related} analyzes related work, and in Section \ref{sec:method} we discuss the methodology used to create the dataset and conduct the experiments. Afterward, Section \ref{sec:discussion} explains the obtained measurements. Finally, the article concludes in Section \ref{sec:conclusions}, highlighting the main takeaways and suggesting future research lines.

\section{Background}
\label{sec:back}
The following sections introduce the relevant tools and metrics used throughout the paper. First, Section \ref{sec:b1} discusses well-known frameworks for vulnerability identification and categorization, as well as the SAST concept. Next, Section \ref{sec:b2} describes the main metrics used in the state of the art for code analysis.

\subsection{Static Code analysis and Weakness Enumeration}
\label{sec:b1}
The Common Weakness Enumeration (CWE) \cite{cwe} by MITRE is the primary framework for identifying and categorizing software and hardware vulnerabilities. It was introduced to standardize the naming and scope conventions of weaknesses. As a result, CWE enables developers, system administrators, and cybersecurity analysts to systematically address the security risks in their organizations by providing a common reference language.

SAST \cite{yang2019towards} is an essential methodology that examines the source code for vulnerabilities without executing it. It enables the early detection of issues such as memory leaks, injection flaws, and dead code. Cppcheck \cite{cppcheck}, as its name implies, focuses on C and C++ code, identifying errors that compromise code quality and security. On the other hand, Bandit \cite{bandit} is another SAST for Python that targets vulnerabilities like improper cryptographic use and hard-coded credentials. Snyk Code \cite{snyk} is a commercial (also available under a freemium model) SAST that supports multiple programming languages and can easily integrate into CI/CD workflows to automate vulnerability detection. Likewise, Semgrep \cite{semgrep} provides an open-source and lightweight alternative, allowing for custom vulnerability rules across diverse programming languages. We consider that the combination of the aforementioned tools and frameworks enables an automated, replicable, comprehensive, and thorough security analysis of code repositories, efficiently supporting the identification of vulnerabilities across almost all major programming languages.

\subsection{Static Code analysis and Code Metrics}
\label{sec:b2}
Recent advancements in static code analysis have led to the development of tools that extract insights from source code by quantifying various features. One such tool is Mozilla's rust-code-analysis \cite{ardito2020rust}, which provides a robust framework for analyzing source code across multiple programming languages, including Rust, JavaScript, Python, C++, and TypeScript. With the use of Abstract Syntax Trees (AST), rust-code-analysis enables an in-depth examination of code structure and syntax, facilitating early error detection and comprehensive metric evaluation.

Code analysis often focuses on three key areas, more precisely:
\begin{itemize}
\item \textbf{Size Metrics:}
Size metrics, such as Physical Lines of Code (PLOC), Logical Lines of Code (LLOC), Comment Lines of Code (CLOC), and Blank Lines of Code, provide a quantitative measure of the code base. These metrics not only reflect the scale of the software but also offer insights into its organization and readability.

\item \textbf{Complexity metrics:}
Complexity metrics, including Cyclomatic Complexity \cite{ebert2016cyclomatic} and Halstead Metrics \cite{halstead1977elements, hariprasad2017software}, quantify the logical complexity of the code. Cyclomatic Complexity assesses the number of linearly independent paths, indicating maintainability, while Halstead Metrics evaluate other aspects such as effort, time, bugs, vocabulary, and difficulty, offering a deeper understanding of the complexity of the code.

\item \textbf{Quality Indicators:}
Quality indicators such as the Maintainability Index (MI) \cite{yenduri2022systematic} and the number of functions can efficiently quantify software maintainability and modularity, highlighting areas for potential improvement. More precisely, MI is a software metric to quantify the ease of maintaining source code by evaluating factors such as cyclomatic complexity, source lines of code (SLOC), and Halstead volume. The higher the values of MI, the more maintainable the code is considered. The original formula \cite{coleman1994using} for MI is given by:

\[
\text{MI}_{\text{original}} = 171 - 5.2 \ln V - 0.23 G - 16.2 \ln L
\]

where $ V $ is the Halstead Volume, $ G $ is the cyclomatic complexity, and $ L $ is the SLOC. The Software Engineering Institute (SEI) introduced a derivative of the MI formula \cite{bray1997c4}, incorporating an additional term for comment lines:

\[
\text{MI}_{\text{SEI}} = 171 - 5.2 \log_2 V - 0.23 G - 16.2 \log_2 L + 50 \sin\left(2.4 \sqrt{C}\right)
\]

where $ C $ denotes the percentage of comment lines (converted to radians). Visual Studio uses a scaled version \cite{mivs} of the original formula, scaling the MI range to $[0, 100]$ and ensuring non-negative values, by using the following formula:

\[
\text{MI}_{\text{VS}} = \max\left[0, 100 \frac{171 - 5.2 \ln V - 0.23 G - 16.2 \ln L}{171}\right]
\]
\end{itemize}

Size metrics, complexity metrics, and quality indicators are intricately correlated, with dependencies that collectively shape the assessment of software systems. Size metrics, such as PLOC and LLOC, directly influence complexity metrics. For instance, an increase in code size often correlates with a rise in Cyclomatic Complexity due to additional control structures and execution paths, which in turn affect maintainability. Halstead Metrics also depend on size-related attributes like the number of operators and operands, linking size directly to cognitive effort and error prediction. Quality indicators like the Maintainability Index depend on both size and complexity metrics, as they are calculated using factors such as Cyclomatic Complexity and Lines of Code.

\section{Related Work}
\label{sec:related}

There are multiple works analyzing malware and how it differs from benign software. However, this article focuses on exploring coding style practices, including a vulnerability analysis, so that an automated pipeline can be leveraged to differentiate among them and outline their distinct structures. In this regard, this article falls outside the scope of author attribution or authorship identification, which is another well-established research line \cite{gray2021identifying}. Note that traditional malware research does not focus on the bugs inherent in malware; instead, static analysis in malware contexts typically involves analyzing malicious binaries to extract static features, e.g., n-grams, imported libraries, to detect them or extract indicators. As such, most prior work on static malware analysis aimed to identify malware behavior or signatures, not to evaluate the security of the malware's code itself.

Barford and Yegneswaran conducted an early qualitative study of four IRC-bot families, demonstrating that, even at a few thousand lines of code, bots already exhibit layered command-and-control logic similar to that found in conventional software projects \cite{Barford07}. Kotov and Massacci analyzed the source code of commercial exploit kits and discovered that these are as large as small open-source programs and contain many CWE-listed vulnerabilities. Their results show that cyber-criminals follow the same basic development routines as ordinary programmers, only with lower security standards \cite{Kotov13}.

Quantitative analyses became possible once larger collections of leaked code were assembled. Calleja \emph{et al.} tracked 30 years of malware evolution, documenting an order-of-magnitude increase in SLOC, function points, and \textsc{COCOMO} effort estimates while observing that malware still lagged contemporary OSS in maintainability \cite{Calleja16}. Their subsequent \textit{MalSource} dataset expanded coverage to 456 samples, uncovering pervasive clone reuse and a gradual convergence toward open-source quality indices \cite{Calleja19}. Frantzeskou \emph{et al.} demonstrated that high-level layout and naming conventions alone are sufficient to attribute authorship across Java and Lisp corpora, hinting at stylometric signatures that may persist in malicious code \cite{Frantzeskou08}; however, Java and Lisp are not the primary programming languages used to write malware.

Although the next group of studies does not target a full, metric-to-metric comparison between malicious and benign projects as we do, they still provide valuable contextual evidence for the feasibility of large-scale source-code research.  Rokon \emph{et al.} designed a crawler that sifted through GitHub and automatically flagged 7,504 repositories as malware, effectively eliminating the long-standing data scarcity problem \cite{Rokon20}. Tereszkowski-Kaminski \emph{et al.} mapped more than 12 million function-level clones across GitHub, StackOverflow, and underground forums, uncovering how malicious snippets propagate between public and clandestine ecosystems rather than evaluating their structural complexity or maintainability \cite{Suarez24}.

A separate Android-focused line of work further highlights the diagnostic power of source-level features, albeit with a primary focus on malware detection.  Cen \emph{et al.} relied on simple API-call frequency vectors extracted from decompiled Java and attained an F1 score of 0.95, underscoring that even coarse-grained features can be highly discriminative \cite{Cen15}.  Tian \emph{et al.} refined this idea by partitioning each app into dependence-based regions \cite{Tian20}. Their proposed heterogeneity-aware model reduced the false-negative rate to 0.35\% on repackaged samples, yet their focus remained on binary classification rather than a systematic comparison of software engineering metrics.

Table \ref{tab:comparison} summarises the main characteristics of the related literature.

\begin{table}[!ht]
\centering
\scriptsize
\caption{Comparison of selected literature. Notation: (\textbf{EK} = Exploit Kit, \textbf{OSS} = Open-Source Software).}
\rowcolors{2}{}{gray!10}
\begin{tabular}{p{1.2cm}p{2cm}p{2cm}p{2.5cm}cp{6cm}}
\toprule
\textbf{Ref.} & \textbf{Malware set} & \textbf{Benign set} & \textbf{Metrics} & \textbf{Vuln.\ scan} & \multicolumn{1}{c}{\textbf{Key highlights}  } \\
\midrule
\cite{Barford07}          & 4 bots              & \xmark             & Architectural taxonomy & \xmark & First qualitative dissection of leaked IRC‐bot source; exposes modular \mbox{C2} layers and reuse of shared libraries, framing bots as engineered products.\\

\cite{Kotov13}              & 30 EKs              & \xmark             & SLOC, CWE density      & \checkmark & Sizes and counts CWEs across exploit-kit families, revealing surprisingly low internal secure-coding standards and common vulnerability hot-spots despite commercial success.\\

\cite{Calleja16}          & 151 leaks           & 9 OSS proj.  & SLOC, FP, CC, COCOMO   & \xmark & 30-year historical trend shows 10× growth in code volume and effort while still lagging contemporary OSS in maintainability.\\

\cite{Calleja19}          & 456 leaks           & 11 OSS proj.  & SLOC, MI, CC, COCOMO   & \xmark & Introduces \textit{MalSource} dataset; quantifies clone-reuse within/between families and charts convergence toward OSS maintainability indices.\\

\cite{Frantzeskou08}  & \xmark                   & 2 Datasets & SCAP stylometry      & \xmark & Demonstrates that comment layout and identifier naming alone can fingerprint programmers across Java and CLisp, paving the way for author-attribution in malware leaks.\\

\cite{Rokon20}              & 7,504 GitHub repos  & 89,871 repos   & SLOC, repo metadata    & \xmark & Presents \textsc{SourceFinder} pipeline that automatically discovers malicious projects at scale, overcoming prior data scarcity for metric studies.\\

\cite{Suarez24}            & 100 repositories  & 100 repositories             & Clone graphs           & \xmark & Maps function-level clones across GitHub, StackOverflow, and others, evidencing cross-ecosystem sharing of malicious templates and tooling.\\

\cite{Cen15}                  & 808 apps          & 173,785 apps     & API-call frequency, permissions & \xmark & Regularised logistic model on decompiled Java achieves F1 = 0.95, proving high discriminative power of simple source-level frequency vectors.\\

\cite{Tian20}                & 1506 + 612 (re-packaged) samples & 5415 Android apps  & Dependence-region API sets & \xmark & Partitions code into heterogeneity regions; combination cuts false-negatives to 0.35\% (false-positives 2.97\%), surfacing malicious payload hidden in benign wrappers.\\

\textbf{This work}           & 658 leaks           & 249 OSS proj. & PLOC, LLOC, CC, Halstead, MI, COCOMO, CWE & \checkmark & First study to use industrial SAST vulnerability enumeration with multi-language size, complexity, and maintainability metrics.\\
\bottomrule
\end{tabular}
\label{tab:comparison}
\end{table}

\section{Methodology}
\label{sec:method}

\begin{figure}[th]
\centering
\includegraphics[width=0.9\linewidth]{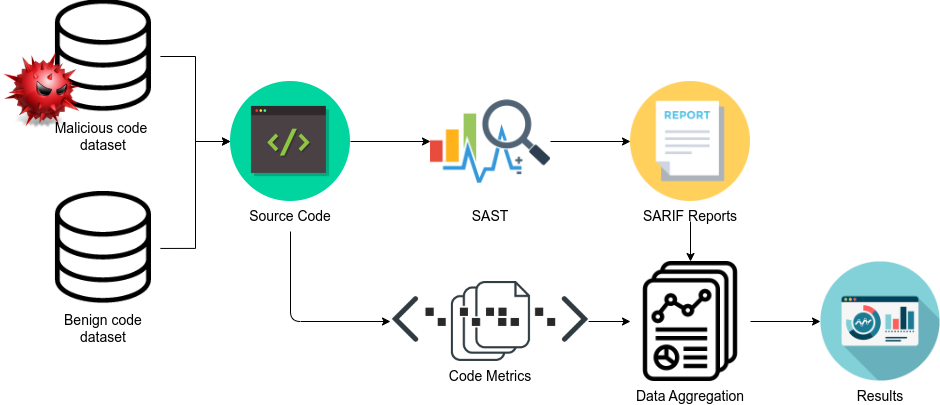}
\caption{Methodology for the analysis pipeline used.}
\label{fig:notes}
\end{figure}

The proposed analysis pipeline is shown in Figure \ref{fig:notes}. The process begins with extracting and arranging the source code to ensure it is structured appropriately for subsequent evaluations. This was necessary because the benign code was properly organized, while the malicious code was not always properly structured, with folders having inconsistent structures, causing parsing issues for the tools. To facilitate the process, in most cases, we resorted to flattening the folder structure so that all malicious source code was transferred to a first-level folder.
Once we organized the source code, it was used as input for SAST specialized tools, see Figure \ref{fig:notes}. These tools examined the code base for potential security vulnerabilities and produced Static Analysis Results Interchange Format (SARIF) \cite{SARIF-v2.1.0} reports, which provide a standardized format for documenting findings, enabling consistent and efficient data handling throughout the workflow.

Next, we analyzed the generated SARIF reports to extract specific security insights. To this end, we identified relevant vulnerabilities assigned to CWEs and the Open Web Application Security Project (OWASP) standards, providing a detailed understanding of the security posture of the code base. In parallel, the source code was analyzed to generate a set of software metrics. These metrics quantified various code attributes, including complexity, maintainability, and other statistical indicators, providing additional context for understanding the code's overall quality and robustness. Finally, the retrieved data, including security weaknesses and code metrics, was combined into a single dataset. This enabled a thorough analysis and accurate interpretation of the data, providing a more comprehensive overview of the quality and security status of the code base.

\section{Experimental Setup}

To conduct our experiments, we have created a benchmark dataset that consists of source code from malicious and benign software. First, we explored VX-Underground \cite{vxunderground} as a source since it is a prominent online repository known for hosting the largest collection of malware samples, source code, and educational materials. More concretely, for the malware dataset, we have used the leaked source code of 658 malware that is available from VX Underground\footnote{\url{https://github.com/vxunderground/MalwareSourceCode}}. The latter is augmented with three additional datasets of benign code that we compiled to allow comparisons between malicious and benign code. The first consists of the source code of the top 100 PyPi packages according to PyPI Stats\footnote{\url{https://pypistats.org/top}}, the second one of the top 100 \texttt{npm} packages according to Socket\footnote{ \url{https://socket.dev/npm/category/popular}}, and then we have 49 open-source cybersecurity projects, including nmap, sqlmap, and zap. From now on, we will refer to them as \texttt{pypi}, \texttt{npm}, and \texttt{cs}, respectively. The metrics for each project are publicly available on Zenodo\footnote{https://zenodo.org/records/20489247}. The selection of the above projects was based on the following rationale. First, we did not include professional projects with lots of authors and precise coding rules and criteria, as it is evident that they differ from malware authors in size and complexity. We also include other benign projects, e.g., repositories with most stars on GitHub\footnote{Using the query: \url{https://github.com/search?q=stars\%3A\>10000\&type=Repositories\&ref=advsearch\&l=\&l=\&s=stars\&o=desc\&p=2}}{ contain many bookmark repositories or large projects (React, the Linux kernel, content management systems, etc.), often backed by large organizations and companies, numerous developers, and applying industry-level practices. However, we wanted something that is smaller and manageable by smaller teams, e.g., libraries, and would not necessarily have industry support. The above led us to consider the cases of \texttt{pypi} and \texttt{npm} libraries for Python and JavaScript, respectively.

Nevertheless, neither case is security-oriented, and we wanted to add a cybersecurity flavor, as some tasks, e.g., network scanning and vulnerability identification, are found in both malicious code and penetration testing tools. Thus, we set as criteria that the code repository (i) is public, (ii) is benign, (iii) is usable, (iv) performs cybersecurity tasks (e.g., network scanning, has payloads, etc.), and (v) has a good reputation (e.g., many stars on GitHub). Our criteria ensure that the selected projects will not be from huge teams and will not have professional or stringent writing rules regarding comments, formatting, etc., as is the case with the Linux kernel. The main statistics of the content of our dataset are described in Table \ref{tbl:dataset}.

For the collection of the code metrics, we selected
Mozilla's rust-code-analysis \cite{ardito2020rust} due to its broad language support and maturity. However, since the tool does not support C\#, Go, or HTML, the metrics were computed on all benign projects but only on a subset of 463 malware codebases. We explicitly acknowledge this coverage limitation, namely the structural and metric-based conclusions drawn in this study represent this specific 70\% subset, ensuring the findings are contextualized within the bounds of the supported languages. 
It is worth noting that for the code-metrics-based behavioural analysis, we applied an outlier filter to the dataset using the standard IQR criterion \cite{tukey1977exploratory}, excluding observations below $Q1 - 1.5 \times IQR$ or above $Q3 + 1.5 \times IQR$, before conducting the experiments discussed below. This step was intended to reduce the influence of extreme values and avoid potential misinterpretation of the results.

To discover the vulnerabilities of the malware, we used Cppcheck, Bandit, Snyk, and Semgrep. This time, no constraint was found, and all dataset samples were analyzed. Finally, to assess the software development effort required for malware creation, we used the Constructive Cost Model (COCOMO). More precisely, we used its Organic variant, which fits small to medium-sized software projects developed by relatively small teams with flexible requirements that we consider match the bulk of the projects in our dataset. The specific parameters and variants of the previous assessments are described in Section \ref{sec:discussion}.

\begin{table}[th]
\centering
\begin{tabular}{cccc}
\toprule
\textbf{Tag}& \textbf{\# projects} & \textbf{Programming language} & \textbf{Availability}  \\
\midrule
\texttt{malware} & 658 & Various & Public leaks\\
\texttt{pypi} & 100 & Python & Github \\
\texttt{npm} & 100 & Javascript & Github \\
\texttt{cs} & 49 & Varius & Github \\
\bottomrule
\end{tabular}
\caption{Overview of our dataset. We use unified tags in lowercase to refer to the subsets for the sake of readability.}
\label{tbl:dataset}
\end{table}
\section{Discussion}
\label{sec:discussion}

In what follows, we describe the outcomes of the analysis, including findings on vulnerability density and code metrics. Given the thorough monitoring of open-source projects and their open nature, we expected the bulk of the findings from static code analysis tools to be false positives; thus, the vulnerability density analysis is performed only on leaked code from malware.

\subsection{Vulnerability Analysis}

When exploring the dataset, we found that malware is mostly coded in low-level languages such as C++ and C (see Table \ref{table:cwe_comp}), which can be attributed to the need for direct system access and aligns with recent statistics \cite{apostolopoulos2025coding}. Other languages, such as PHP and HTML, showed lower vulnerability frequencies, possibly due to their limited use in malware or sophisticated obfuscation.

\begin{table*}[th]
\centering
\setlength{\tabcolsep}{15pt} 
\begin{tabular}{|l|r|l|r|}
\hline
\textbf{File Extension} & \textbf{Count} & \textbf{File Extension} & \textbf{Count} \\
\hline
cpp & 22735 & html & 4657 \\
\hline
c & 15165 & php & 3596 \\
\hline
h & 2571 & js & 1077 \\
\hline
py & 1012 & cs & 796 \\
\hline
java & 253 & cxx & 224 \\
\hline
aspx & 165 & go & 158 \\
\hline
cc & 131 & hpp & 126 \\
\hline

\end{tabular}
\caption{Dataset file extension composition of CWE findings.}
\label{table:cwe_comp}
\end{table*}

Interestingly, the identified OWASP \cite{owasp} rules significantly deviate from the OWASP Top 10 (Table \ref{tbl:trigered_rules}), which serves as a guideline for identifying the most critical security risks in web applications. This imbalance highlights the differing priorities of malware authors' intentions. Furthermore, distinct clusters of common weaknesses can be observed across malware source code samples, underscoring shared behavioral patterns and development methodologies among threat actors as seen in Figure \ref{fig:malware_cwe}.

\begin{table}[ht]
\centering
\small
\begin{subtable}[t]{\textwidth}
\centering
\rowcolors{2}{}{gray!10}
\begin{tabular}{ccp{5in}}
\toprule
\textbf{CWE} & \textbf{Count} & \textbf{CWE title}\\
\midrule
398 & 20033 & 7PK - Code Quality\\
353 & 4054 & Missing Support for Integrity Check\\
563 & 4025 & Assignment to Variable without Use\\
561 & 3454 & Dead Code\\
686 & 2212 & Function Call With Incorrect Argument Type\\
676 & 1490 & Use of Potentially Dangerous Function\\
89  & 1374 & Improper Neutralization of Special Elements used in an SQL Command (SQL Injection')\\ 457 & 1206 & Use of Uninitialized Variable \\ 79  & 1162 & Improper Neutralization of Input During Web Page Generation (Cross-site Scripting')\\
628 & 961 & Function Call with Incorrectly Specified Arguments \\
401 & 854 & Missing Release of Memory after Effective Lifetime \\
570 & 760 & Expression is Always False\\
571 & 703 & Expression is Always True\\
190 & 665 & Integer Overflow or Wraparound\\
78 &  601 & Improper Neutralization of Special Elements used in an OS Command (`OS Command Injection')\\
319 & 562 & Cleartext Transmission of Sensitive Information\\
476 & 540 & NULL Pointer Dereference\\
916 & 535 & Use of Password Hash With Insufficient Computational Effort\\
23 &  520 & Relative Path Traversal\\
685 &  480 & Function Call With Incorrect Number of Arguments\\
\bottomrule
\end{tabular}
\caption{Top 20 Most Common CWEs Categories.}

\end{subtable}
\begin{subtable}[t]{0.32\textwidth}
\centering
\begin{tabular}{cc}
\toprule
\textbf{OWASP 2017} & \textbf{Count} \\
\midrule
A01 & 868 \\
A03 & 622 \\
A07 & 489 \\
A06 & 257 \\
A05 & 126 \\
A08 & 72  \\
A04 & 12  \\
A02 & 11  \\
\bottomrule
\end{tabular}
\caption{OWASP-2017 Issues}
\end{subtable}
~
\begin{subtable}[t]{0.32\textwidth}
\centering
\begin{tabular}{cc}
\toprule
\textbf{OWASP 2021} & \textbf{Count} \\
\midrule
A08 & 4271 \\
A03 & 1959 \\
A02 & 663  \\
A10 & 286  \\
A05 & 269  \\
A01 & 199  \\
A07 & 62   \\
A04 & 11   \\
\bottomrule
\end{tabular}
\caption{OWASP-2021 Issues}
\end{subtable}
\caption{Top vulnerabilities by CWE, OWASP-2017, and OWASP-2021 in the malware dataset}
\label{tbl:trigered_rules}
\end{table}

\begin{figure}[!ht]
\centering
\includegraphics[width=0.9\linewidth]{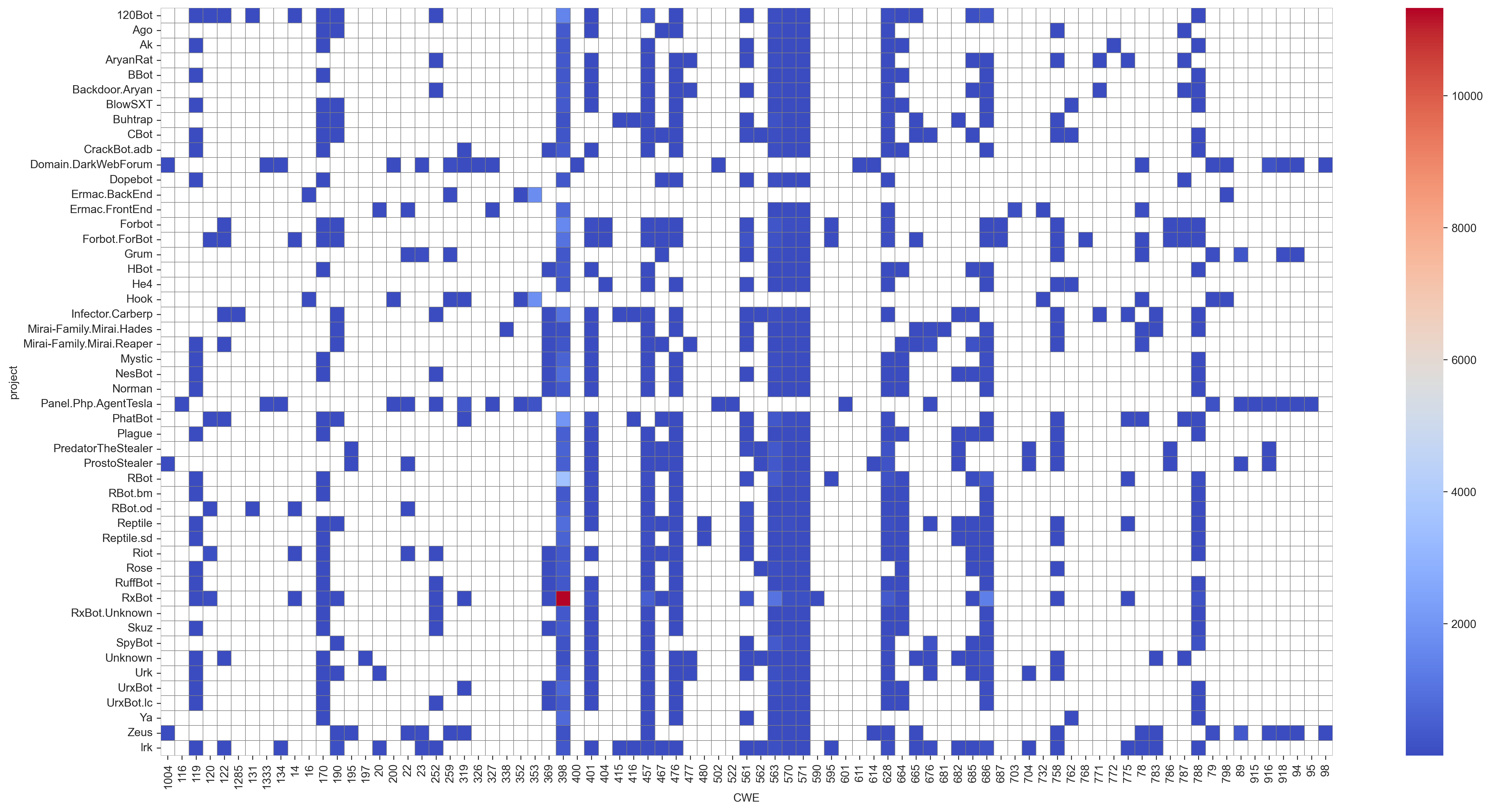}
\caption{ Distribution of CWE findings across the 50 malicious code bases with the highest CWE counts.}
\label{fig:malware_cwe}
\end{figure}

One of the key findings regarding code security issues is the prevalence of CWE-398 (Poor Code Quality), CWE-563 (Assignment to Variable without Use), and CWE-561 (Dead Code) \cite{cwe}, which emphasize that malware developers prioritize operational objectives over code maintainability and quality. Indeed, as illustrated in Table \ref{tbl:trigered_rules}, among the top 20 CWEs, many are code quality-related CWEs. Beyond the three aforementioned, one can observe CWE-457 (Use of Uninitialized Variable) far too often, as well as CWE-628 and CWE-685, which point to erroneous function calls. Moreover, CWE-570 and CWE-571 signify wrong use of expressions, as some expressions are always False and True, which can be linked to dead code in the sense of, e.g., not executing a specific code segment. Of particular interest is the use of insecure channels to transmit information (CWE-319), which allows interception and manipulation of the traffic, e.g., via the command and control (C2) server (see Listing \ref{lst:unencrypted}). However, disabling SSL/TLS certificate validation, as seen in Listing \ref{lst:tlscert}, may be a deliberate choice in some cases, as malware authors would not use proper certificates to avoid attribution. Given that this flaw allows man-in-the-middle interception/modification, enabling remote monitoring and detection of infected hosts, we consider this an unexpected finding. On the contrary, the lack of proper input sanitization (CWE-89, CWE-79, and CWE-78) is expected, as malware typically requires little user interaction and can hardly impact the host infection or its persistence (e.g., see Listing \ref{lst:SQLinjection}). In the same context, CWE-676, CWE-190, CWE-916, and CWE-23, while they would significantly impact a typical application, may be less likely to affect the malware's primary operational goal in some samples. Hence, their presence and number imply pure negligence on the part of the malware authors.

\begin{table}[!th]
\centering
\caption{Detailed CWE Statistical Analysis for Malware and Benign Samples}
\label{tab:cwe_analysis_final}
\footnotesize
\begin{tabular}{l p{4.5cm} r r c c l} 
\toprule
CWE & Description & Mal \% & Ben \% & 95\% CI & p & Direction \\ 
\midrule
CWE-398 & Code Quality & 52.80 & 27.50 & [1.90, 1.95] & $< 0.001$ & Malware $\uparrow$ \\
CWE-563 & Unused Variable & 8.70 & 5.10 & [1.65, 1.75] & $< 0.001$ & Malware $\uparrow$ \\
CWE-686 & Argument Mismatch & 5.40 & 1.20 & [4.28, 4.69] & $< 0.001$ & Malware $\uparrow$ \\
CWE-353 & Missing Integrity Check & 4.10 & 0.00 & [952.80, 7751.13] & $< 0.001$ & Malware $\uparrow$ \\
CWE-561 & Dead Code & 3.50 & 5.80 & [0.58, 0.63] & $< 0.001$ & Benign $\uparrow$ \\
CWE-457 & Uninitialized Variable & 2.80 & 0.40 & [6.79, 7.89] & $< 0.001$ & Malware $\uparrow$ \\
CWE-628 & Argument Type Mismatch & 2.70 & 3.90 & [0.66, 0.72] & $< 0.001$ & Benign $\uparrow$ \\
CWE-676 & Dangerous Function & 1.50 & 0.00 & [470.11, 11558.64] & $< 0.001$ & Malware $\uparrow$ \\
CWE-570 & Always False & 1.50 & 2.60 & [0.54, 0.61] & $< 0.001$ & Benign $\uparrow$ \\
CWE-571 & Always True & 1.50 & 4.20 & [0.33, 0.37] & $< 0.001$ & Benign $\uparrow$ \\
CWE-89 & SQL Injection & 1.40 & 0.20 & [7.24, 9.05] & $< 0.001$ & Malware $\uparrow$ \\
CWE-79 & Cross-site Scripting & 1.20 & 0.10 & [8.12, 10.49] & $< 0.001$ & Malware $\uparrow$ \\
CWE-401 & Memory Leak & 1.20 & 0.20 & [4.65, 5.72] & $< 0.001$ & Malware $\uparrow$ \\
CWE-476 & NULL Pointer Deref & 0.90 & 1.30 & [0.63, 0.73] & $< 0.001$ & Benign $\uparrow$ \\
CWE-788 & Buffer After End & 0.80 & 0.00 & [21.91, 35.62] & $< 0.001$ & Malware $\uparrow$ \\
CWE-685 & Wrong Argument Count & 0.70 & 0.00 & [110.07, 464.00] & $< 0.001$ & Malware $\uparrow$ \\
CWE-190 & Integer Overflow & 0.70 & 0.60 & [0.99, 1.19] & 0.072 & Equal \\
CWE-78 & OS Command Injection & 0.60 & 0.90 & [0.59, 0.71] & $< 0.001$ & Benign $\uparrow$ \\
CWE-319 & Cleartext Transmission & 0.60 & 0.00 & [33.73, 73.52] & $< 0.001$ & Malware $\uparrow$ \\
CWE-916 & Weak Password Hash & 0.50 & 0.10 & [5.82, 8.14] & $< 0.001$ & Malware $\uparrow$ \\
\bottomrule
\end{tabular}
\end{table}

The outcomes seen in Table \ref{tab:cwe_analysis_final} further support this comparison by showing that the most prevalent CWE categories are distributed differently across malicious and benign code. In particular, malware exhibits markedly higher rates of broad code-quality issues CWE-398, unused variables CWE-563, argument mismatches CWE-686, and missing integrity checks CWE-353, all of which are consistent with rapid and less robust development practices. By contrast, benign samples show relatively higher proportions for issues such as dead code CWE-561, argument type mismatches CWE-628, always-false conditions CWE-570, and always-true conditions CWE-571. Overall, the distribution of the top CWE categories indicates that malware and benign software differ not only in the volume of findings but also in the types of weaknesses most commonly observed in each group.

It is interesting to note here that of the 658 malware code samples, 89.82\% (n=591) had at least one CWE, and 112 different CWE categories were reported in total. This suggests that the analyzed malware code bases exhibit a wide range of security issues. Given that a CWE could be reported multiple times, we depicted the distribution of the number of CWEs per malware in Figure \ref{fig:cwe_histogram}. As it can be observed, most malware has a low number of CWEs, yet a small number of them exhibit a high number of CWEs (i.e., with several of them reaching above 60 different CWEs).

\begin{figure}[ht]
\centering
\includegraphics[width=0.75\linewidth]{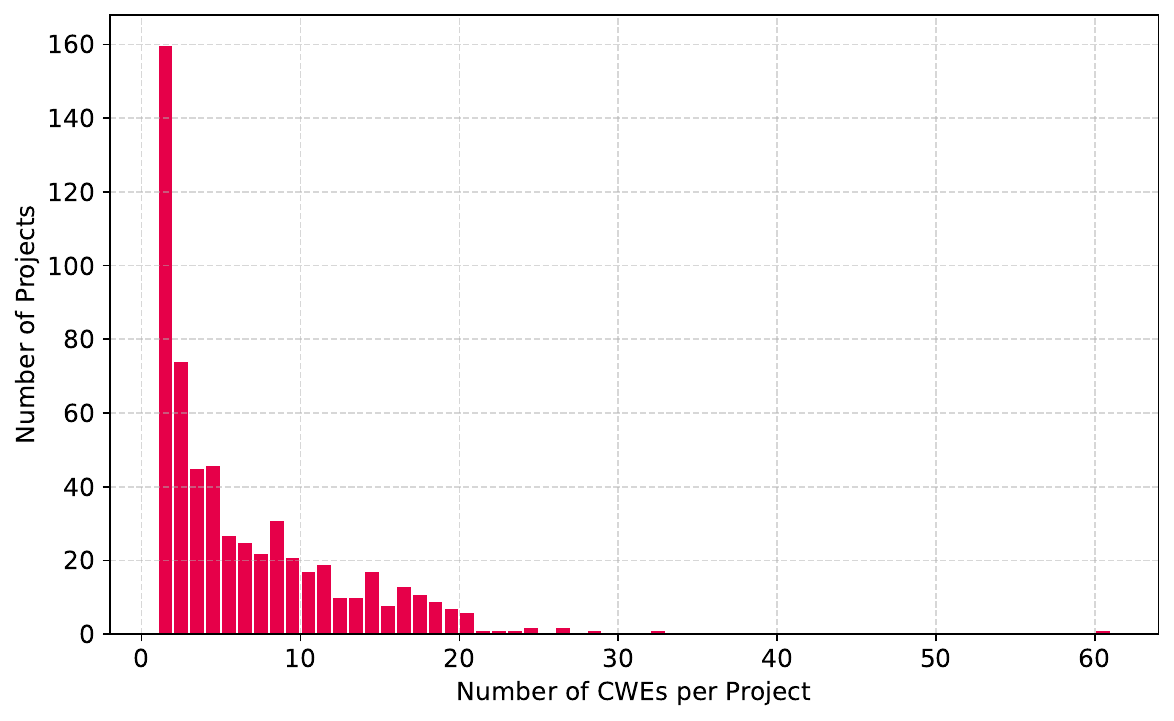}
\caption{Histogram of CWE counts per malware project.}
\label{fig:cwe_histogram}
\end{figure}

\begin{lstlisting}[language=java,caption={Unencrypted socket used by Dendroid.},label={lst:unencrypted}]
protected String doInBackground(String... params) {
for (long stop=System.nanoTime()+TimeUnit.MILLISECONDS.toNanos(Integer.parseInt(j));stop>System.nanoTime();) {
try {
String target = i;
Socket net = new Socket(target, 80);
sendRawLine("GET / HTTP/1.1", net);
sendRawLine("Host: " + target, net);
sendRawLine("Connection: close", net);
}
catch(UnknownHostException e)
{}
catch(IOException e)
{}
}
return "Executed";
}
\end{lstlisting}

\begin{lstlisting}[language=Golang,caption={Disabling TLS/SSL certificate verification in moobot.},label={lst:tlscert}]
var i int
startedTime := time.Now().Unix()
tr := &http.Transport {
TLSClientConfig: &tls.Config{InsecureSkipVerify: true},
}
rand.Seed(time.Now().UTC().UnixNano())
targetRand := randomToken(target)
\end{lstlisting}

\begin{lstlisting}[language=php,caption={Snippet of vulnerable code to SQL injection in Anubis.},label={lst:SQLinjection}]
foreach($imeis_ as $imei){
if($imei !="" ){
$command_ = "|command=Send_GO_SMS|number=$numb|text=$msg::";
$statement = $connection3->prepare("insert into commands (IMEI,command)value ('$imei','$command_')");
$statement->execute(array($imei,$command_));
}
}
\end{lstlisting}

\subsection{Code Metrics}

\begin{figure}[th]
\centering
\begin{subfigure}[hb]{0.49\textwidth}
\includegraphics[width=\textwidth]{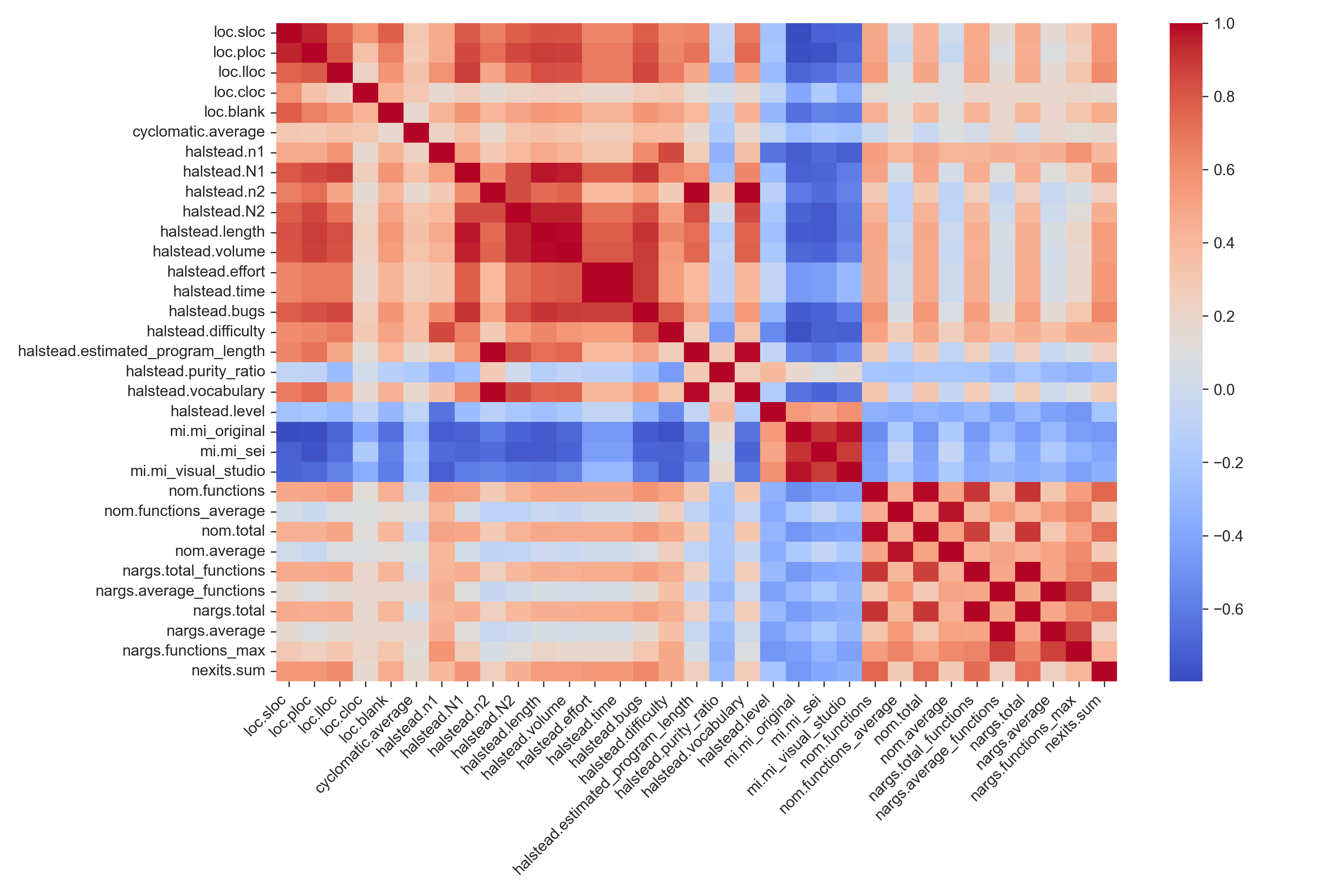}
\caption{Malicious code}
\end{subfigure}~
\begin{subfigure}[hb]{0.49\textwidth}
\includegraphics[width=\textwidth]{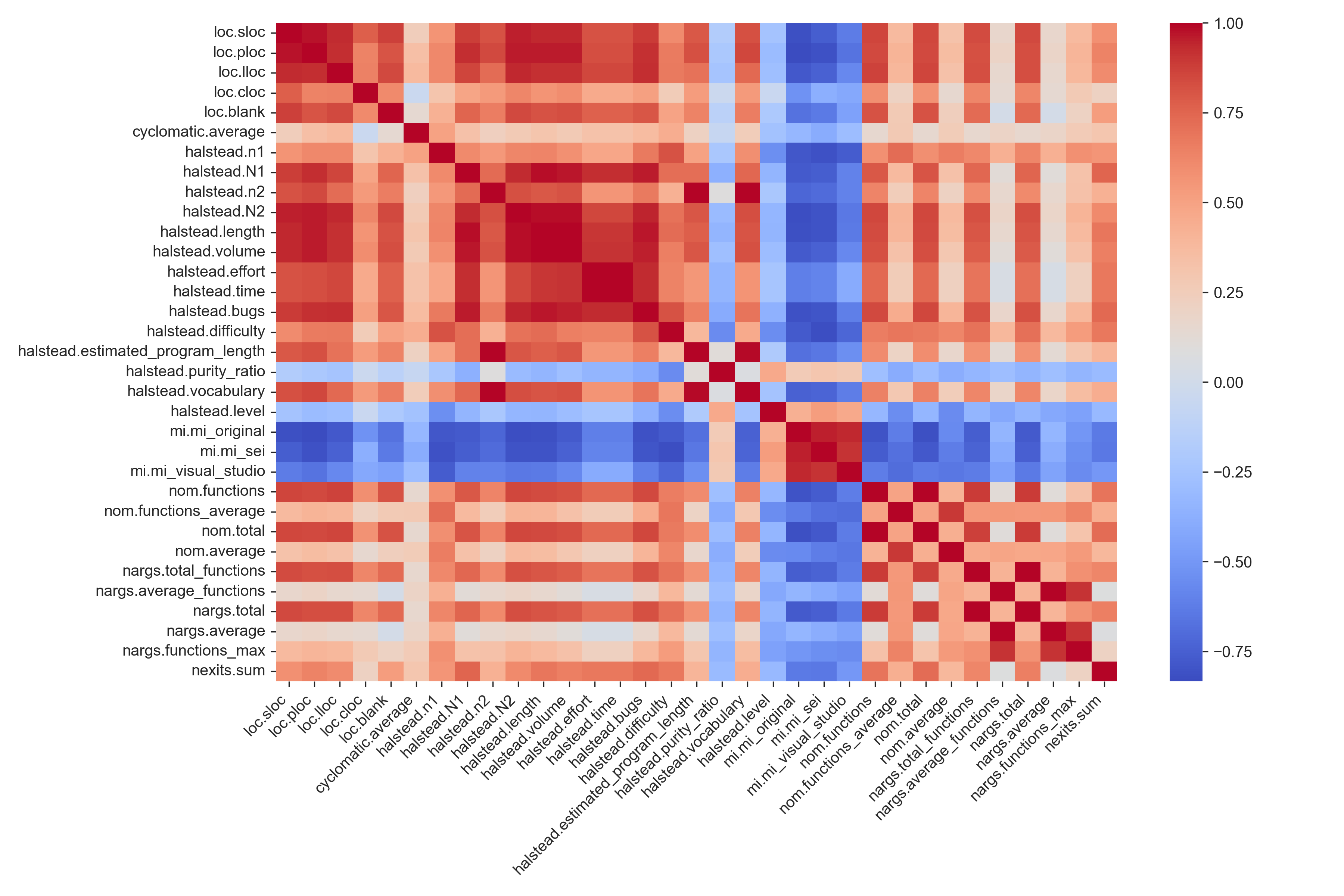}
\caption{Code from top 100 \texttt{pypi} projects.}
\end{subfigure}

\begin{subfigure}[hb]{0.49\textwidth}
    \includegraphics[width=\textwidth]{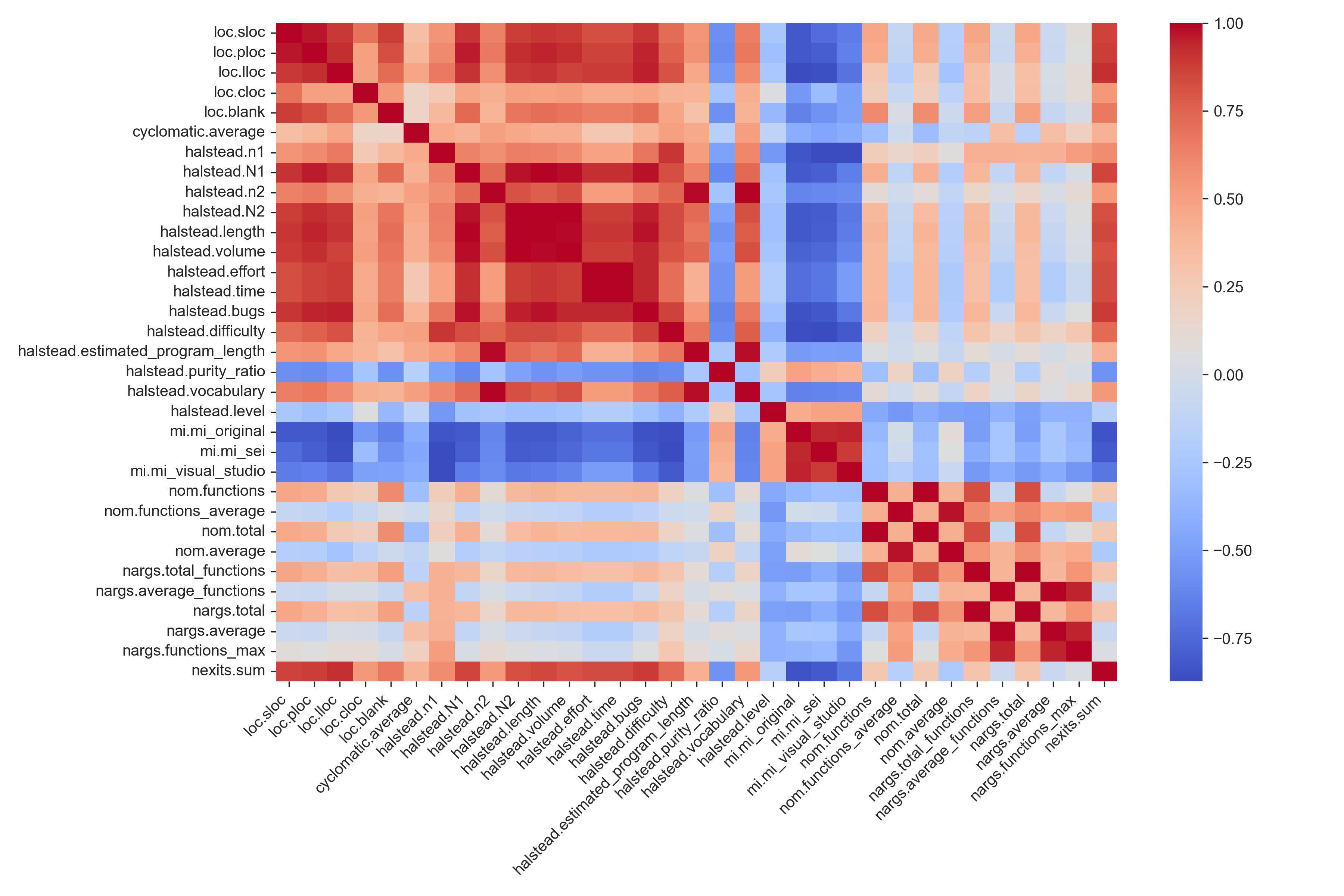}
    \caption{Code from \texttt{cs} projects.}
\end{subfigure}~
\begin{subfigure}[hb]{0.49\textwidth}
    \includegraphics[width=\textwidth]{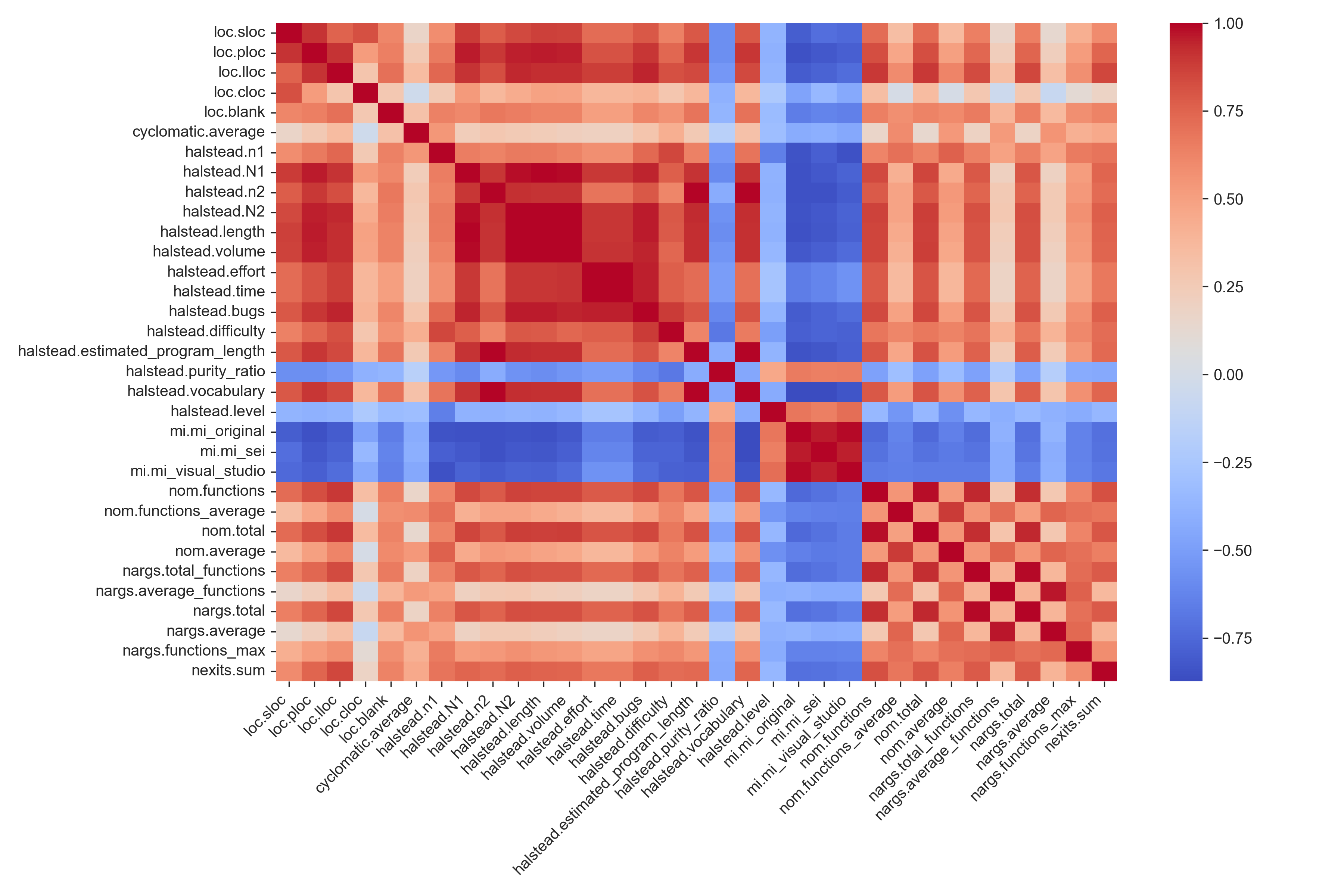}
    \caption{Code from top 100 \texttt{npm} projects.}
\end{subfigure}

\caption{Heat map illustrating the correlation matrix for the dataset features across the four categories.}
\label{fig:feature_heatmap}

\end{figure}

First, we extracted a set of code metrics from the four subsets of our dataset, which are detailed in Table \ref{tbl:metrics}, using Mozilla's rust-code-analysis. While many of these metrics are well-known and widely used, we kindly point the reader to \cite{ardito2020rust} for more details.
Next, we attempted to determine how they correlate to extract some potential insights. To do the latter, we generated the correlation matrices for each subset and depicted them in Figure \ref{fig:feature_heatmap}. While some of them are highly correlated, e.g., the number of source code lines with total source code lines, and Halstead length with volume, other metrics are not correlated with each other, e.g., maintenance-related metrics with other metrics. Moreover, there are no apparent correlations that appear only in specific segments of the dataset, e.g., Halstead metrics with SLOC and PLOC are highly correlated in \texttt{cs} projects compared to other cases.

\begin{figure}[th]
\centering
\includegraphics[width=0.9\linewidth]{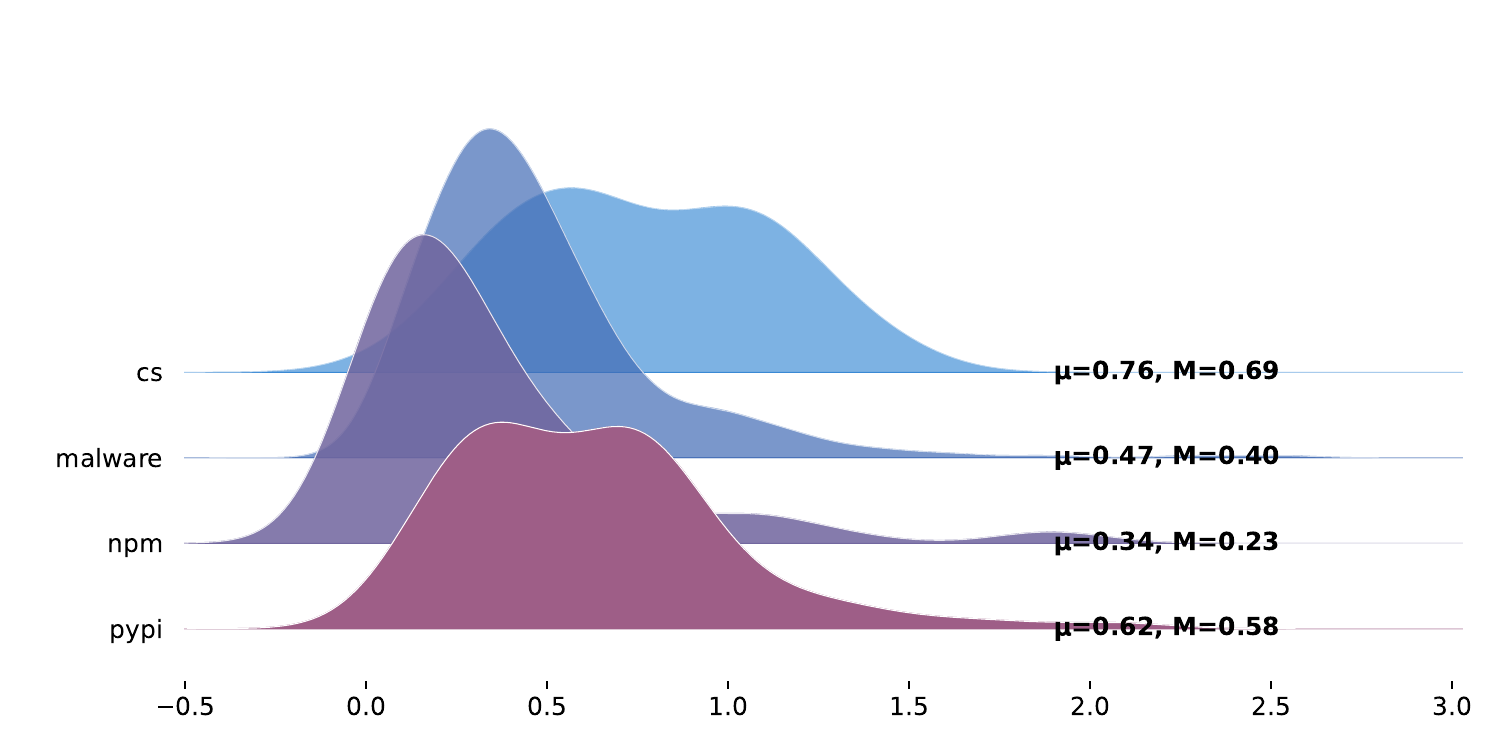}
\caption{COCOMO effort distribution by software category.}
\label{fig:COCOMO_effort_distribution}
\end{figure}

We applied the COCOMO model to estimate the effort in person-months based on thousands of source lines of code (KLOC), with the Organic model's constants set to $a=2.4$ and $b=1.05$. In our analysis, malware projects showed notably low COCOMO effort estimates, highlighting their typically compact nature, as seen in Figure \ref{fig:COCOMO_effort_distribution}. Specifically, the median effort required for malware samples was approximately 0.40 person-months, with a mean effort slightly higher at about 0.47 person-months. The range of effort estimates for malware was relatively narrow, with a minimum of around 0.003 and a maximum near 2.53 person-months, indicating that most malware tends to involve relatively minimal development effort. On the contrary, benign software (\texttt{cs}, \texttt{pypi}, and \texttt{npm} datasets) generally presented higher COCOMO effort estimates, reflecting their larger code bases and more comprehensive features.

\begin{figure}[th]
\centering
\includegraphics[width=0.9\linewidth]{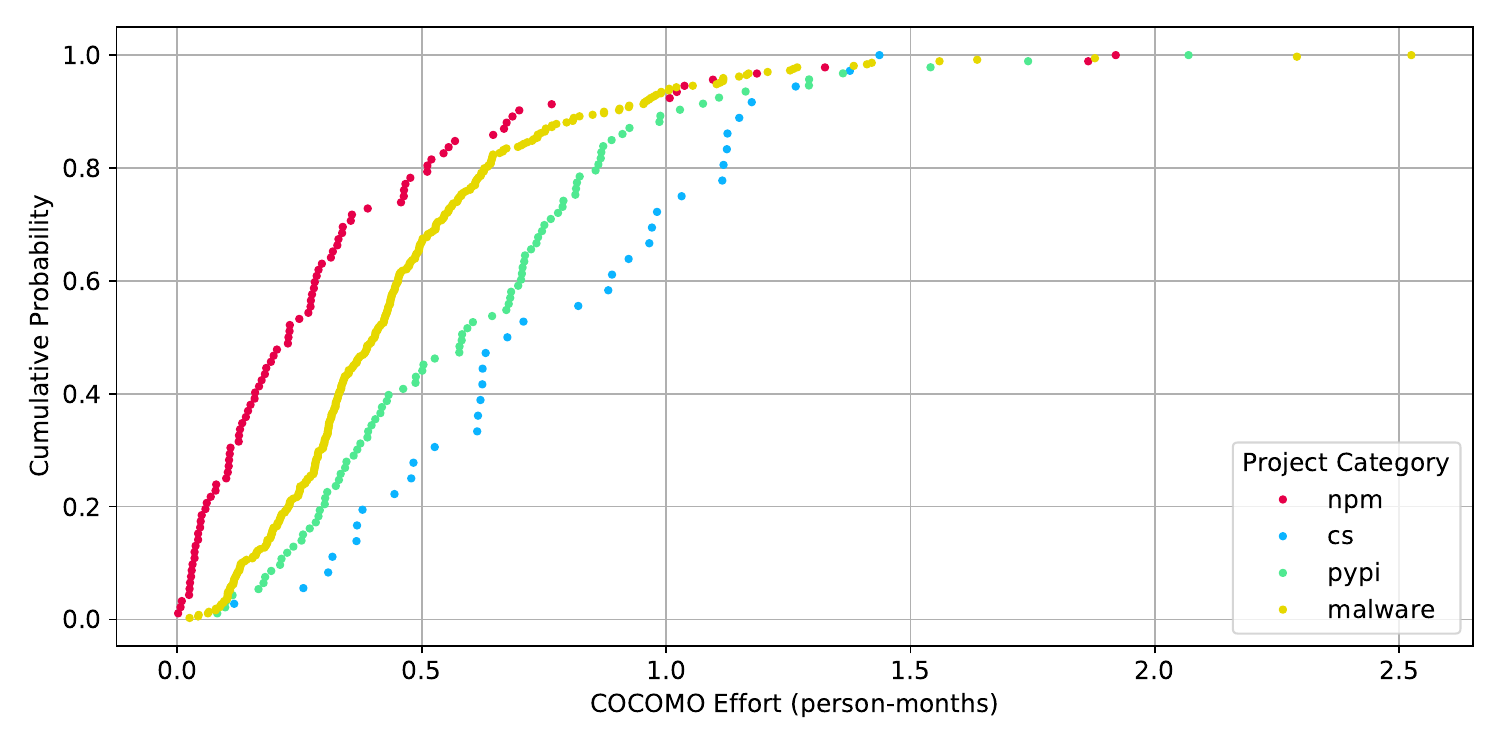}
\caption{COCOMO effort by code category CDF.}
\label{fig:COCOMO_CDF}
\end{figure}

Moreover, as indicated in Figure \ref{fig:COCOMO_CDF}, most of the \texttt{npm} code bases have a significantly lower COCOMO effort than the rest of the datasets. Nevertheless, while 80\% of the projects in the entire dataset seem to have an effort of around one person-month or below, the malware that requires more than one person-month effort significantly outnumber the other categories,

\begin{figure}[th]
\centering
\includegraphics[width=0.9\linewidth]{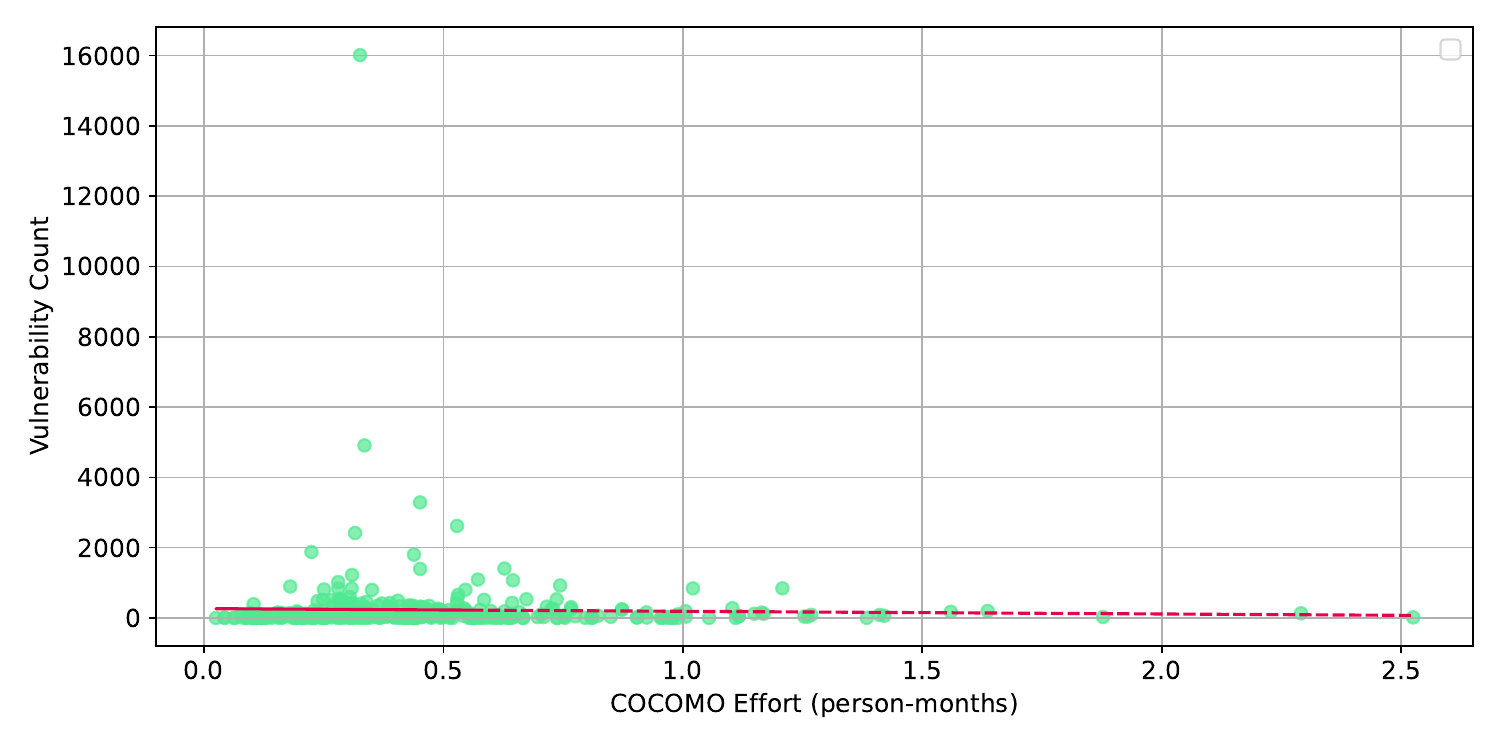}
\caption{Correlation between COCOMO effort and vulnerability count. The red line denotes the correlation trend between the effort and the number of vulnerabilities.}
\label{fig:COCOMO_vulnerability}
\end{figure}

When trying to correlate the COCOMO index with the vulnerability count (Figure \ref{fig:COCOMO_vulnerability}), we observe that malware built with little effort has many reported CWEs, some of which are on the scale of thousands. While having so many vulnerabilities may seem absurd, it is practically a result of the way that some CWEs are being reported by SAST, for instance, one CWE per file, with poor code quality.
On the contrary, more mature malware code bases or malware requiring higher effort to be developed have a substantially lower number of vulnerabilities. As a result, the trend line is almost constant with a low decline, and thus, we can claim that the more effort spent on coding, the lower the number of vulnerabilities found. Moreover, we observed that malware typically involves small, quick-to-develop code bases, aligning with the attacker's imperative for swift development cycles and limited investment in long-term code maintainability or extensive documentation. This modest estimated effort does not equate to reduced operational effectiveness; instead, it highlights a common strategy in malware development—minimalism and efficiency. Interestingly, outliers with higher COCOMO values indicate more complex malware or heavily obfuscated code, which artificially inflate the source lines of code, subsequently increasing effort estimates.

In general, malware code bases tend to be relatively small compared to many benign projects. In our dataset, malware projects have a median ($M$) source lines of code (SLOC) around $183.33$ and mean $(\mu) \approx 208.51$, whereas benign projects from \texttt{pypi} and \texttt{cs} are larger ($\mu=272.61, M=259.74$ for \texttt{pypi} and $\mu=334.62, M=305.84$ for \texttt{cs}) and smaller for \texttt{npm} packages ($\mu=150.59, M=106.50$), see Figure \ref{fig:distr_sloc}. Alongside size, documentation is a key differentiator: malware code is typically sparsely commented, see Figure \ref{fig:distr_ratio_comments}. The median malware project had only about $\mu=21\%$ of its lines as comments (often just a few comment lines as $M=0.13$), whereas benign code (especially \texttt{cs}/\texttt{pypi} projects) showed a significantly higher comment density, ($\mu=0.30, M=0.26$ for \texttt{pypi}, $\mu=0.60, M=0.27$ for \texttt{npm}, and $\mu=0.42, M=0.29$ for \texttt{cs}). We attribute this disparity to the concept that attackers include few explanatory comments to avoid aiding code analysis or their peers easily forking their code without their aid.

When comparing control-flow complexity, malware shows slightly higher cyclomatic complexity on average per function than benign code. In our dataset, the median average cyclomatic complexity per function is about 2.58 for malware, versus 2.01 to 2.66 for benign projects, see Figure \ref{fig:distr_avg_cyclo}. This suggests that malware functions often contain more branches or decision points. Legitimate \texttt{npm} packages, in particular, have very low complexity (many functions with cyclomatic complexity 1 to 2). Despite this, the overall complexity differences are not extreme, i.e., benign software can also have complex functions, especially in larger \texttt{pypi} or \texttt{cs} projects. Interestingly, the MI metrics do not portray malware as poorly written, as seen in Figure \ref{fig:distr_vs_mi_vs} for the Visual Studio Maintainability Index, even though the corresponding CWEs appear more frequently in the static analysis. On the contrary, malware had MI values on par with or even slightly better than some benign code. For example, malware code in our dataset exhibited a median Visual Studio maintainability index of around 33.10, similar to \texttt{npm} projects ($\approx 36.38$) and slightly lower than \texttt{pypi} ($\approx 29.15$). Given the slight variations in the datasets, we can partly attribute the higher MI to the fact that many malware samples are small, with lower SLOC and simpler module structures, which boosts maintainability scores. Nonetheless, when comparing with the MI (Figure \ref{fig:distr_vs_mi}), the quality issues in malware code become more apparent, with many samples displaying low or negative values, indicating poor maintainability due to issues such as high complexity, excessive code length, or insufficient documentation.

The Halstead metrics, which quantify operators, operands, and complexity, reinforce the above observations. Benign \texttt{npm} packages, being the smallest code bases, have the lowest Halstead values, e.g., low program length and volume (i.e., the median volume is approximately half that of others) and low calculated effort. Larger benign projects (\texttt{pypi}, \texttt{cs}) and malware have higher lengths and volumes on average. For instance, the median Halstead volume for malware is $\approx 5.6k$, slightly lower than \texttt{pypi} projects ($\approx 6.7k$) and below \texttt{cs} ($\approx 10.2k$), whereas \texttt{npm} packages are much lower ($\approx 3.4k$), see Figure \ref{fig:distr_halstead_volume}.

The Halstead difficulty (the ratio of unique to total operators/operands) is in a similar range for malware and large benign code (around 23-35), indicating that the intrinsic complexity of the vocabulary is comparable, see Figure \ref{fig:distr_halstead_difficulty}. One notable difference is in the extremes, as malware exhibits some outliers with enormously high Halstead metrics. A low number of malware samples have Halstead volume orders of one magnitude larger (into the hundreds of thousands or millions), and correspondingly huge `effort' values, far exceeding anything seen in the benign subsets, as seen in Figure \ref{fig:distr_halstead_effort}. These cases likely stem from highly obfuscated or auto-generated malware code that uses a vast number of operations, inflating the Halstead counts. The Halstead-derived bug count metric (an estimate of inherent defects) hovers around $\approx 1.1$ malware and $\approx  1.4$ for \texttt{pypi}, with CS exhibiting notably higher counts ($\approx  2.3$) due to its larger code bases, while simple \texttt{npm} modules score lower ($\approx  0.6$), as seen in Figure \ref{fig:distr_halstead_bugs}. Overall, Halstead measures correlate with code size and complexity; malware that is small and straightforward has low Halstead values similar to small benign code, while more complex malware overlaps with the range of larger benign projects. Only the outliers, considering the Halstead complexity, hint at malicious obfuscation, as benign software rarely exhibits such extreme values.
Malware code also exhibits distinctive patterns in its structure, including the organization of functions and other units. Malware projects typically define only a few functions (often just 1 to 5 functions in total, with many having a single main routine). In our dataset (see Figure \ref{fig:distr_nom.functions}), the median malware sample had $\approx 3.13$ functions, compared to $\approx 12.69$ in \texttt{pypi} projects and $\approx 9.13$ in \texttt{cs} projects. \texttt{npm} packages also had a low function count (median $\approx 3.58$), reflecting their minimal scope. Moreover, malware functions tend to be very simple in interface as they accept very few parameters, see Figure \ref{fig:distr_avg_arg_fun}. The average number of arguments per function in malware is less than one, as many functions do not take any arguments, and the maximum number of parameters in a typical malware project is only about two. In contrast, legitimate software often defines more complex APIs. \texttt{pypi} and \texttt{cs} projects frequently have at least one function with $\approx 3-4$ parameters, so their median maximum arguments are on the scale of  $2.5$ vs $\approx 1.5$ for malware, as seen in Figure \ref{fig:distr_nargs.functions_max}. The latter indicates that malware functions are self-contained and not designed for reusability or flexibility, whereas benign code (especially in libraries) uses more parameters to handle various inputs. Another notable remark is the use of classes and closures, as seen in Figures \ref{fig:distr_npm.classes} and \ref{fig:distr_nom.closures}, respectively. The vast majority of malware samples had no classes or interfaces defined (nearly 0\% OOP usage in malware), suggesting an imperative or script-based style. Similarly, several benign programs did not use classes (especially considering the sample of small projects), but a big portion (40\% of the sampled \texttt{cs} projects) of the larger benign projects did, reflecting typical software engineering practices, Figure \ref{fig:distr_npm.classes}. Closures (e.g., inner functions or lambda/function expressions) were almost absent in malware, yet quite common in benign code written in JavaScript or Python, see Figure \ref{fig:distr_nom.closures}. \texttt{npm} packages, in particular, made heavy use of closures, e.g., callback functions and immediately-invoked function expressions, something that was rarely observed in malware code. This results in a higher number of closure constructs in benign code, for example, the median \texttt{npm} project still featured non-zero closures ($ \approx 0.7$), and the mean was over one, whereas the malware median was zero. Thus, we can state that malware is often structured as a flat script or a small set of functions, with little modularity. On the contrary, legitimate projects tend to have more abstractions, more functions (and often smaller functions), occasional classes, and a greater use of structured programming constructs, demonstrating a radically different coding mentality.

\begin{figure}[!th]
\centering
\begin{subfigure}[hb]{0.49\textwidth}
\includegraphics[width=\textwidth]{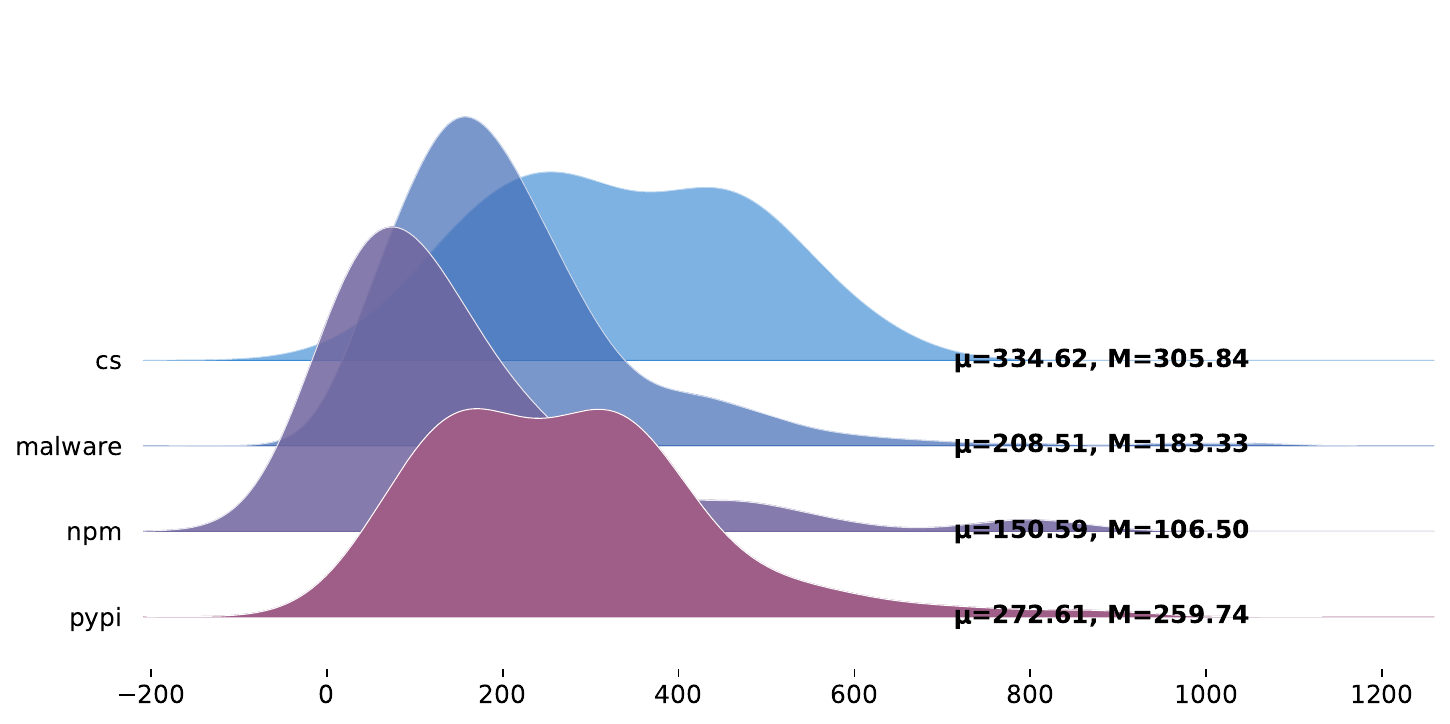}
\caption{Distribution of source lines of code (SLOC).}
\label{fig:distr_sloc}
\end{subfigure}\hfill
\begin{subfigure}[hb]{0.49\textwidth}    \centering
\includegraphics[width=\textwidth]{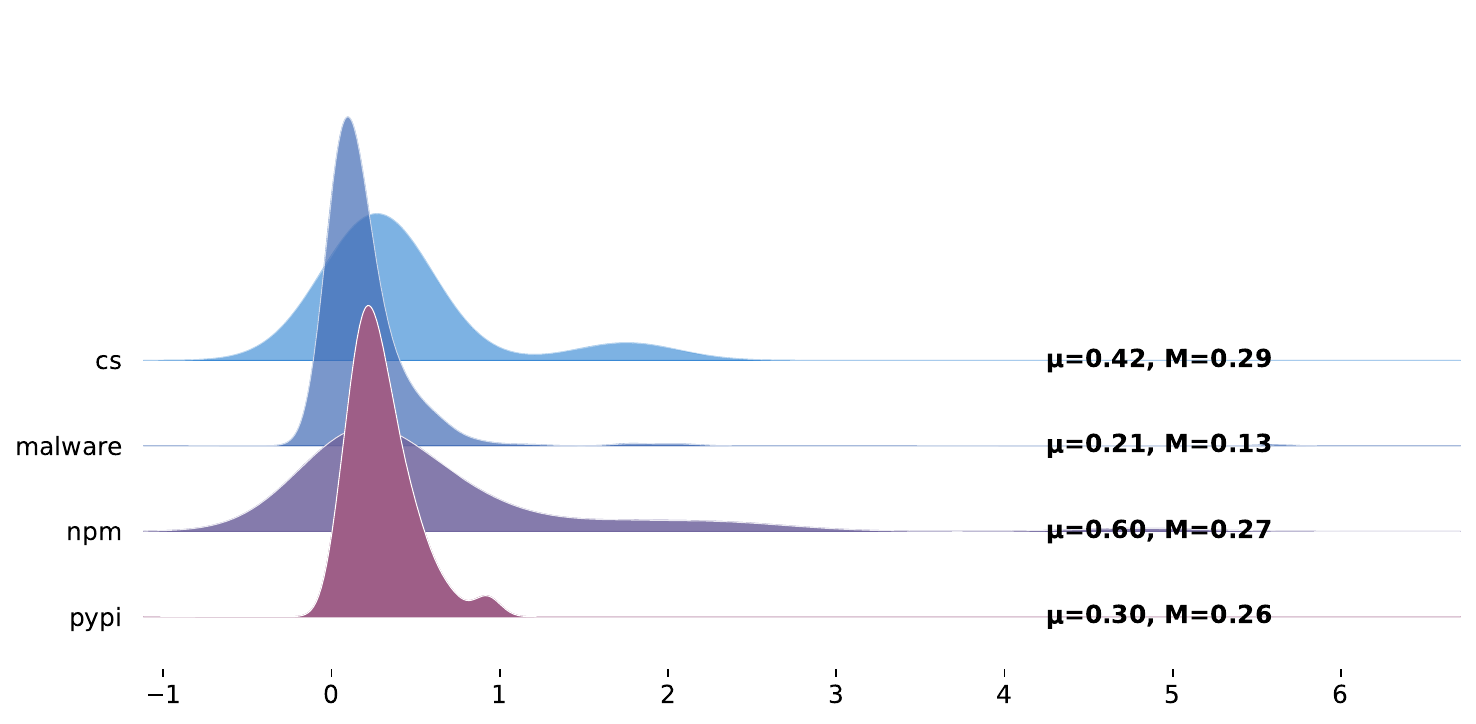}
\caption{Distribution of ratio of comment lines of code to total lines of code (CLOC/PLOC).}
\label{fig:distr_ratio_comments}
\end{subfigure}

\begin{subfigure}[hb]{0.49\textwidth}
    \includegraphics[width=\textwidth]{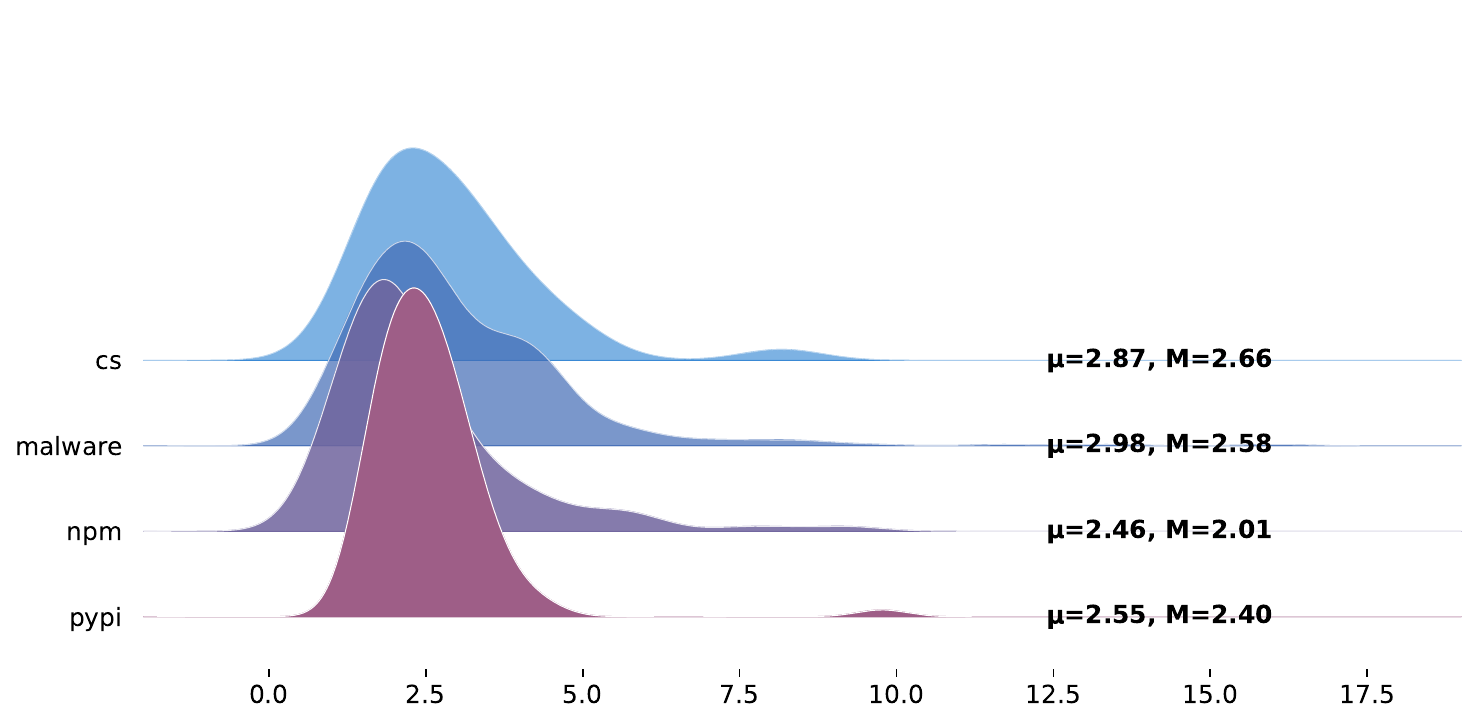}
    \caption{Distribution of average cyclomatic complexity per function.}
    \label{fig:distr_avg_cyclo}
\end{subfigure}\hfill
\begin{subfigure}[hb]{0.49\textwidth}    \centering
\includegraphics[width=\textwidth]{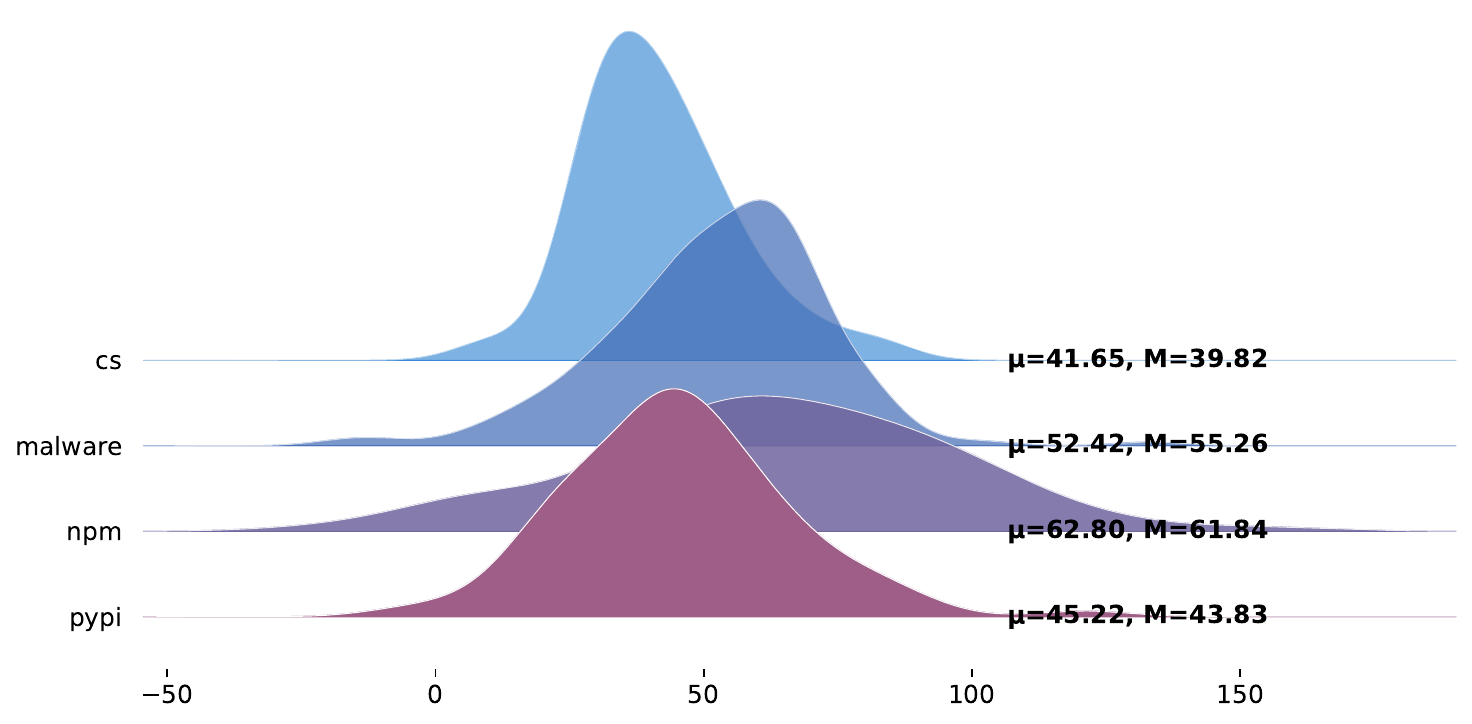}
\caption{Distribution of the Maintainability Index.}
\label{fig:distr_vs_mi}
\end{subfigure}

\begin{subfigure}[hb]{0.49\textwidth}    \centering
\includegraphics[width=\textwidth]{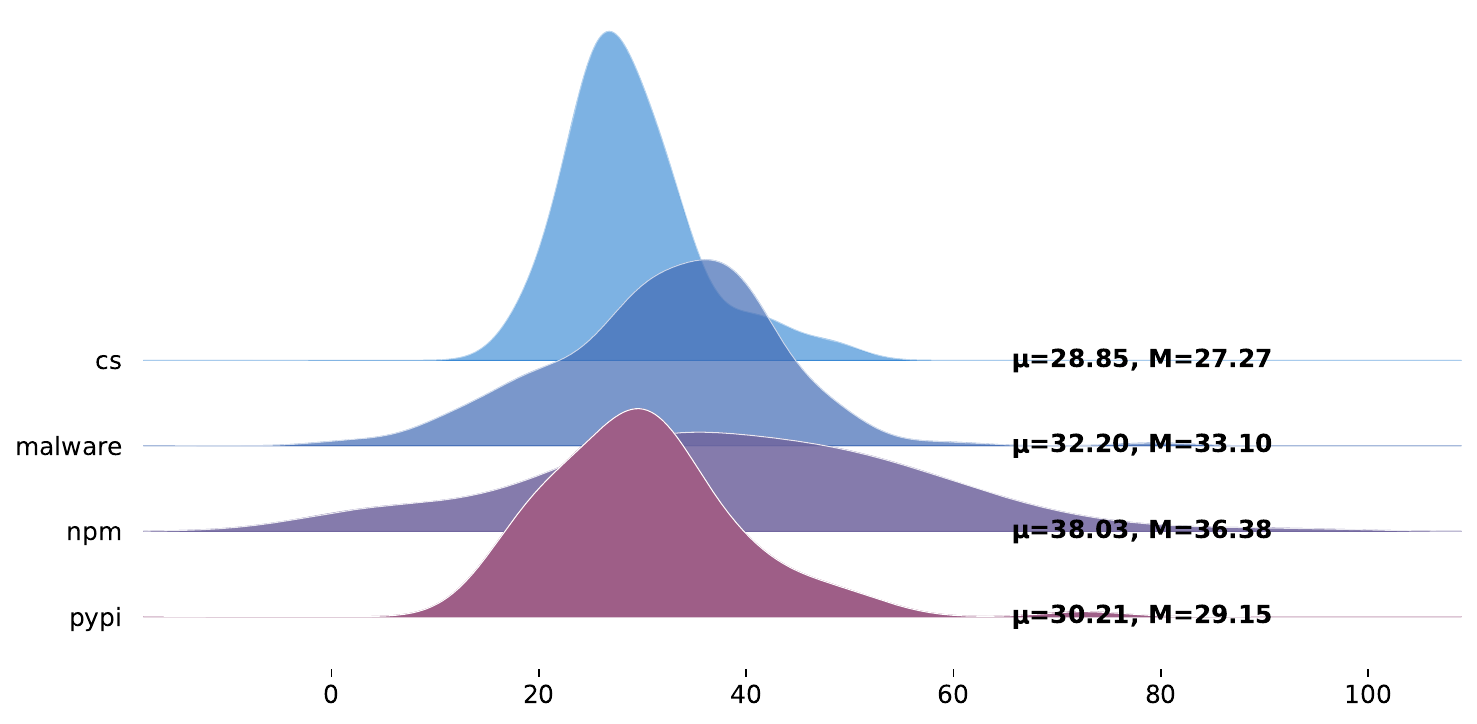}
\caption{Distribution of the Visual Studio-style Maintainability Index.}
\label{fig:distr_vs_mi_vs}
\end{subfigure}\hfill
\begin{subfigure}[hb]{0.49\textwidth}
    \includegraphics[width=\textwidth]{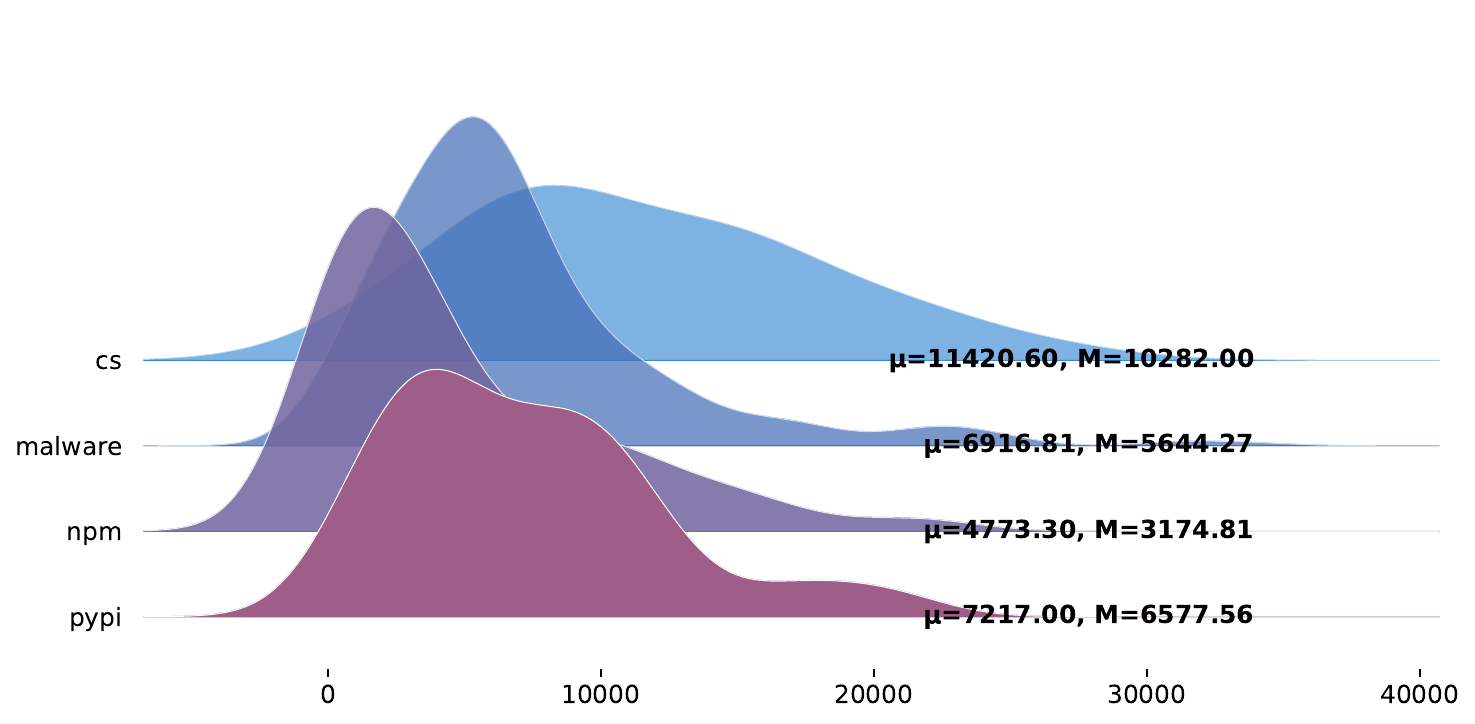}
    \caption{Distribution of Halstead Volume.}
    \label{fig:distr_halstead_volume}
\end{subfigure}

\begin{subfigure}[hb]{0.49\textwidth}    \centering
\includegraphics[width=\textwidth]{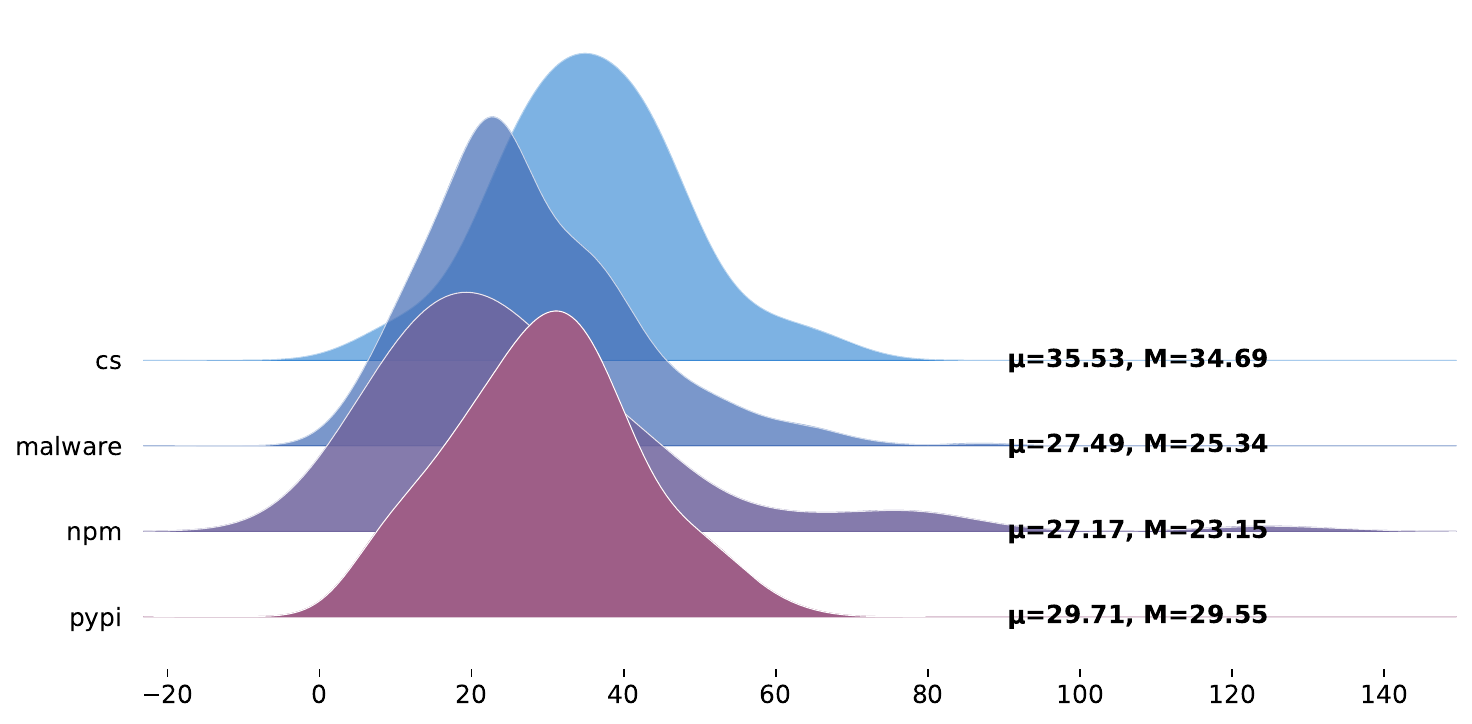}
\caption{Distribution of Halstead difficulty.}
\label{fig:distr_halstead_difficulty}
\end{subfigure}\hfill
\begin{subfigure}[hb]{0.49\textwidth}
    \includegraphics[width=\textwidth]{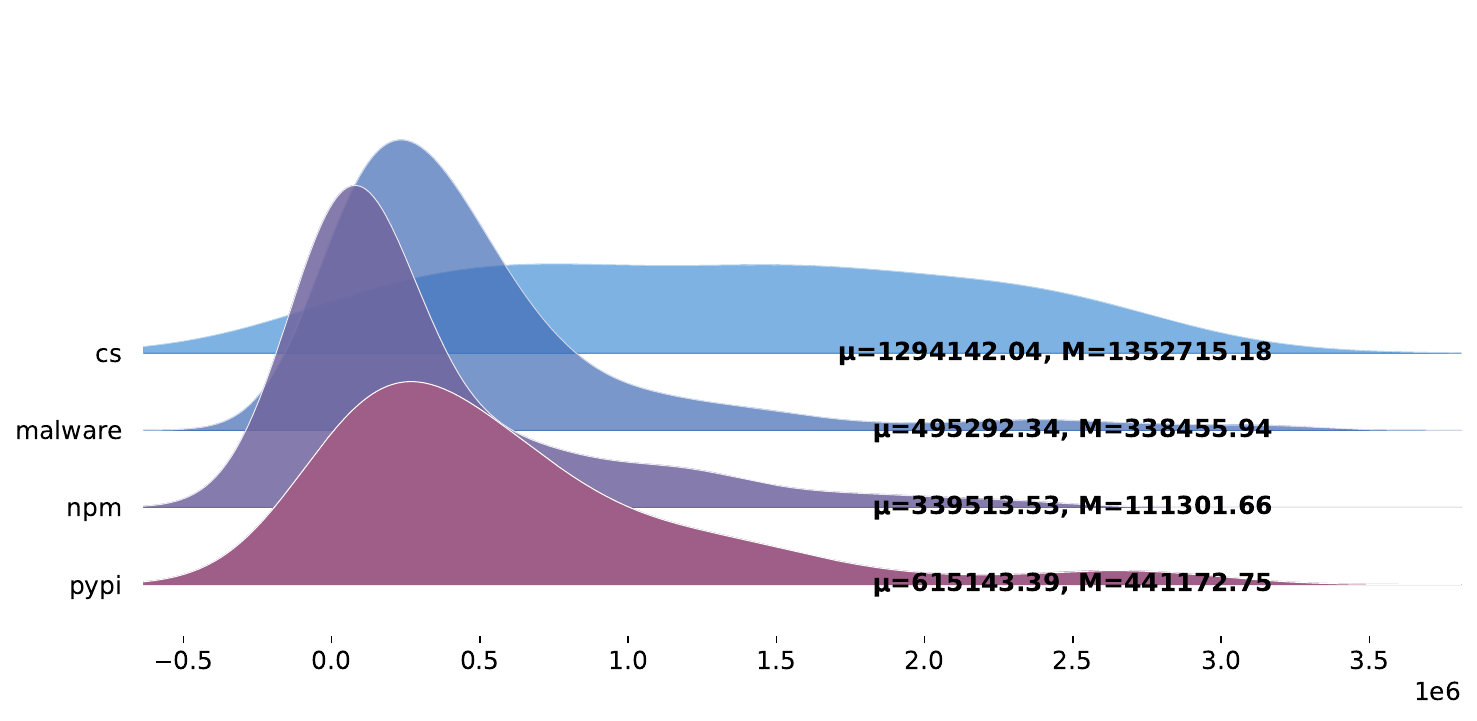}
    \caption{Distribution of Halstead effort.}
    \label{fig:distr_halstead_effort}
\end{subfigure}\hfill
\caption{Distribution of metrics per code category. Notation: $\mu$ denotes mean, and  M denotes median.}

\end{figure}

\begin{figure}[!ht]
\begin{subfigure}[hb]{0.49\textwidth}
\includegraphics[width=\textwidth]{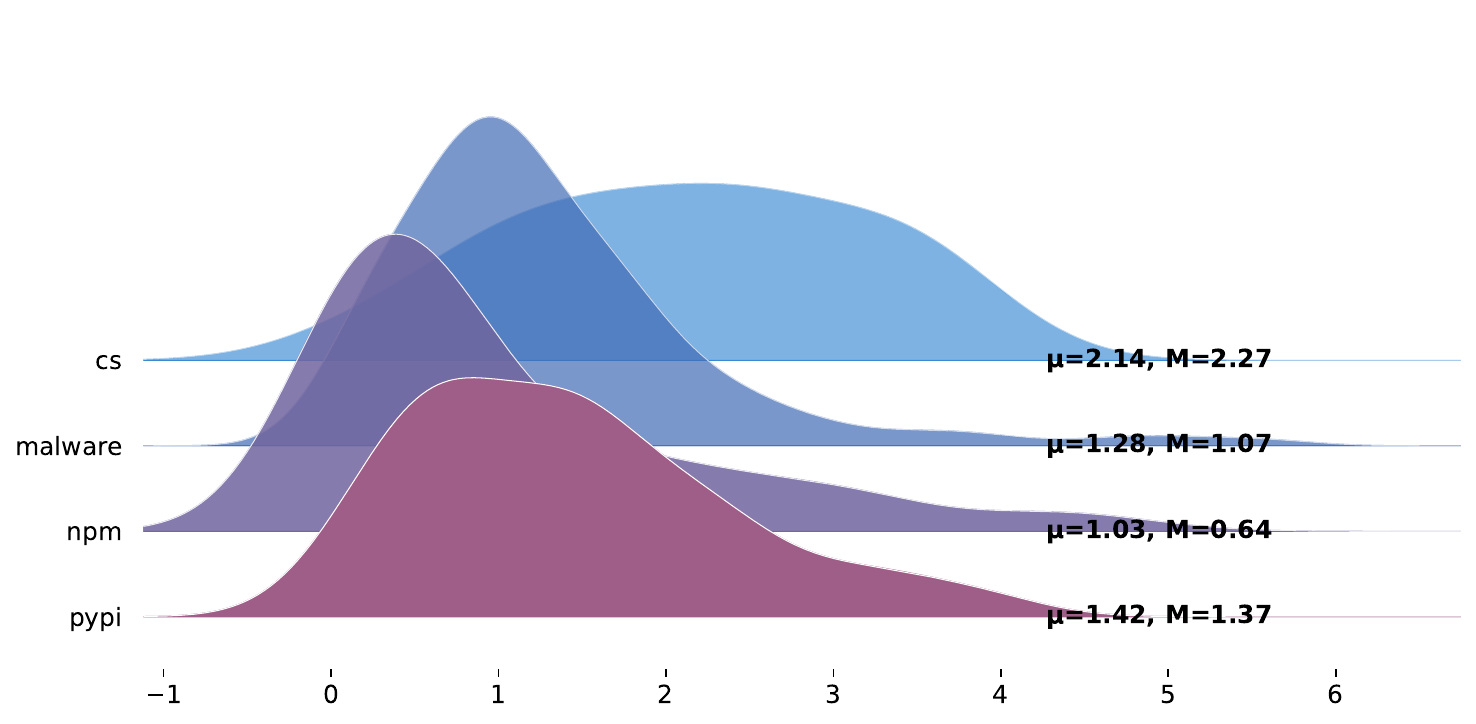}
\caption{Distribution of Halstead bugs.}
\label{fig:distr_halstead_bugs}
\end{subfigure}\hfill
\begin{subfigure}[hb]{0.49\textwidth}    \centering
\includegraphics[width=\textwidth]{images/distr_kde_metrics.loc.sloc.pdf}
\caption{Distribution of source lines of code (SLOC).}
\label{fig:distr_sloc2}
\end{subfigure}

\begin{subfigure}[hb]{0.49\textwidth}
    \includegraphics[width=\textwidth]{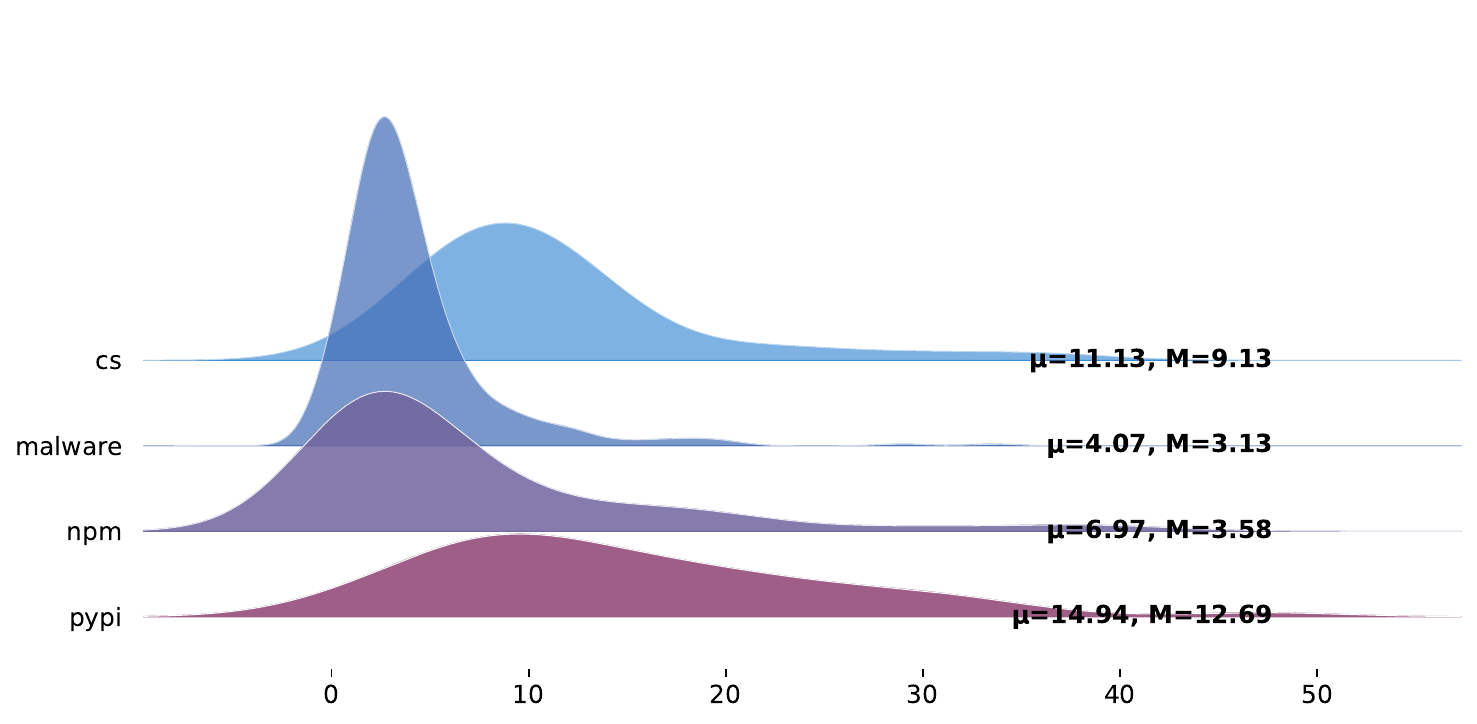}
    \caption{Distribution of number of functions.}
    \label{fig:distr_nom.functions}
\end{subfigure}\hfill
\begin{subfigure}[hb]{0.49\textwidth}    \centering
\includegraphics[width=\textwidth]{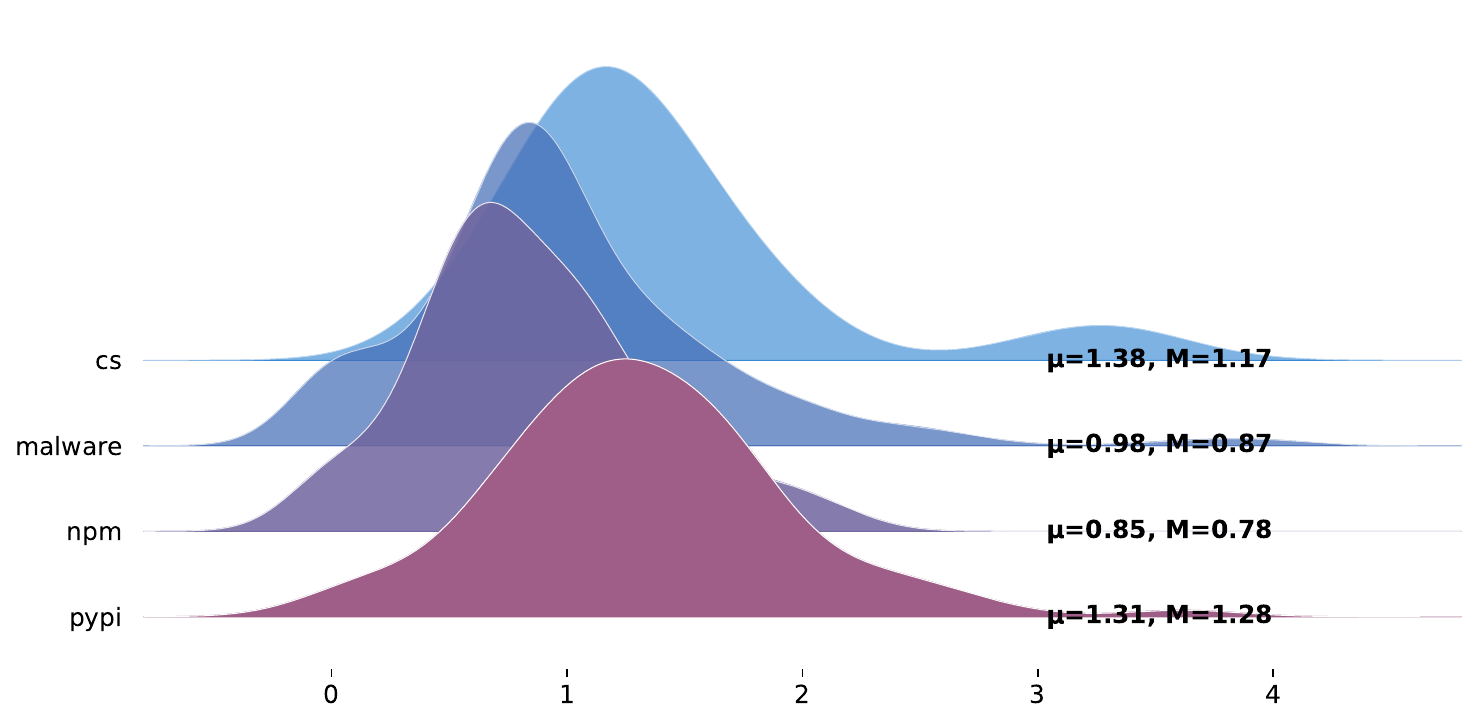}
\caption{Distribution of average arguments per function.}
\label{fig:distr_avg_arg_fun}
\end{subfigure}

\begin{subfigure}[hb]{0.49\textwidth}
    \includegraphics[width=\textwidth]{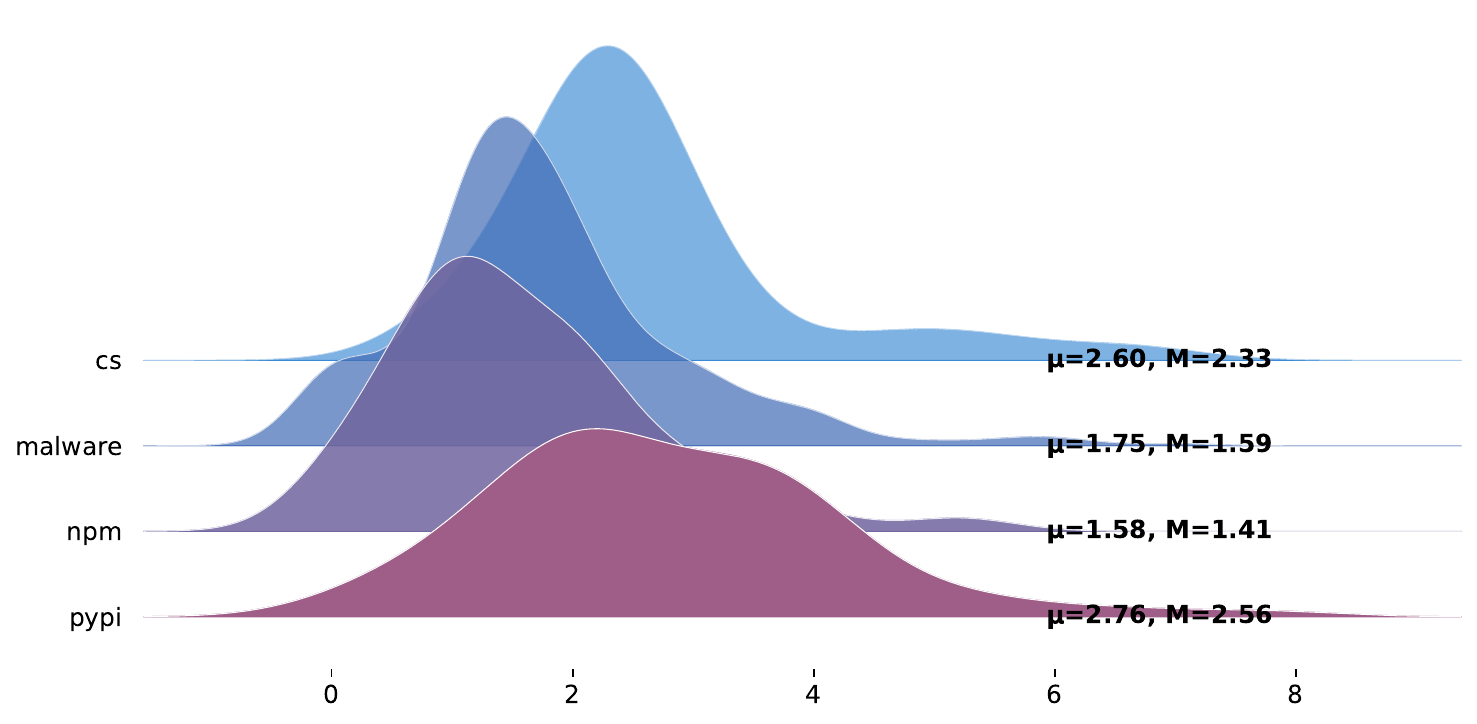}
    \caption{Distribution of Halstead bugs.}
    \label{fig:distr_functions_max.pdf}
\end{subfigure}\hfill
\begin{subfigure}[hb]{0.49\textwidth}    \centering
\includegraphics[width=\textwidth]{images/distr_kde_metrics.nargs.functions_max.pdf}
\caption{Distribution of maximum number of function arguments.}
\label{fig:distr_nargs.functions_max}
\end{subfigure}

\begin{subfigure}[hb]{0.49\textwidth}
    \includegraphics[width=\textwidth]{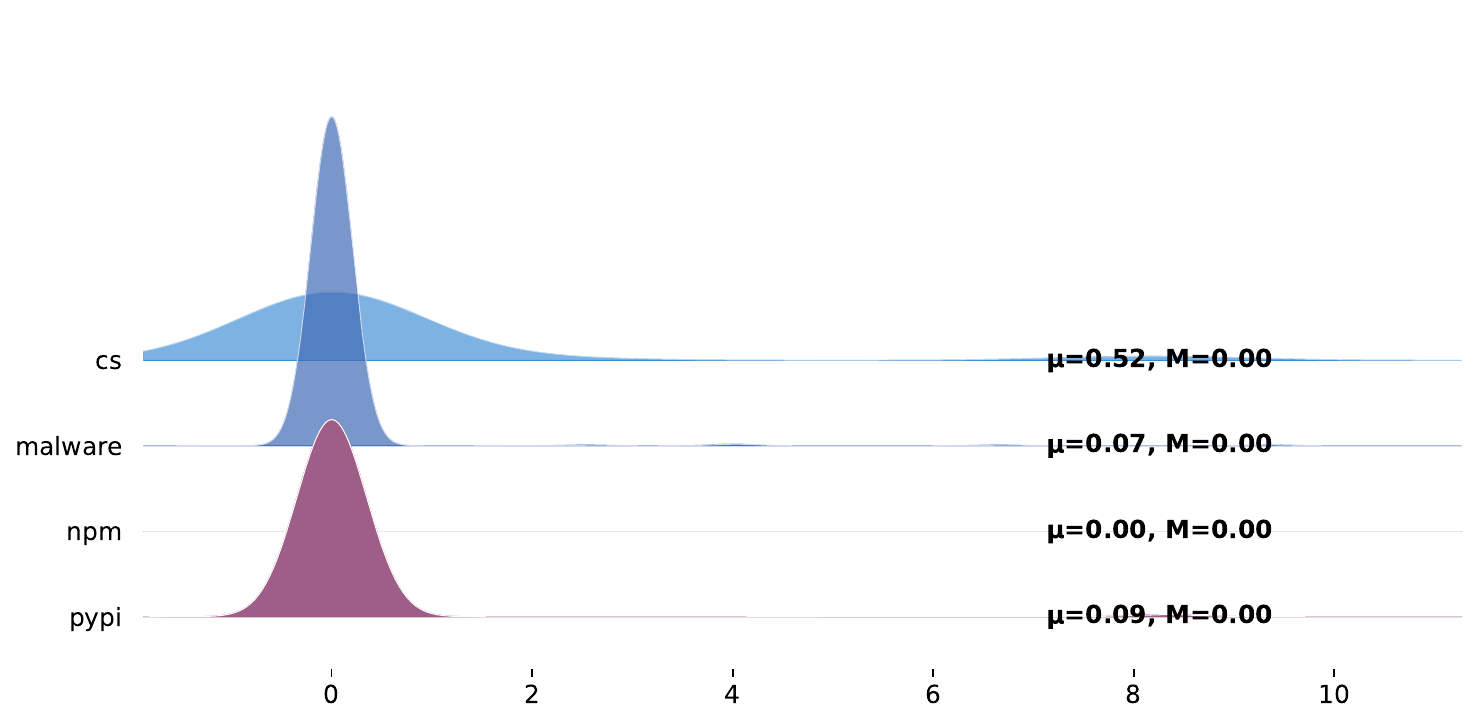}
    \caption{Distribution of classes.}
    \label{fig:distr_npm.classes}
\end{subfigure}
\hfill
\begin{subfigure}[hb]{0.49\textwidth}
    \includegraphics[width=\textwidth]{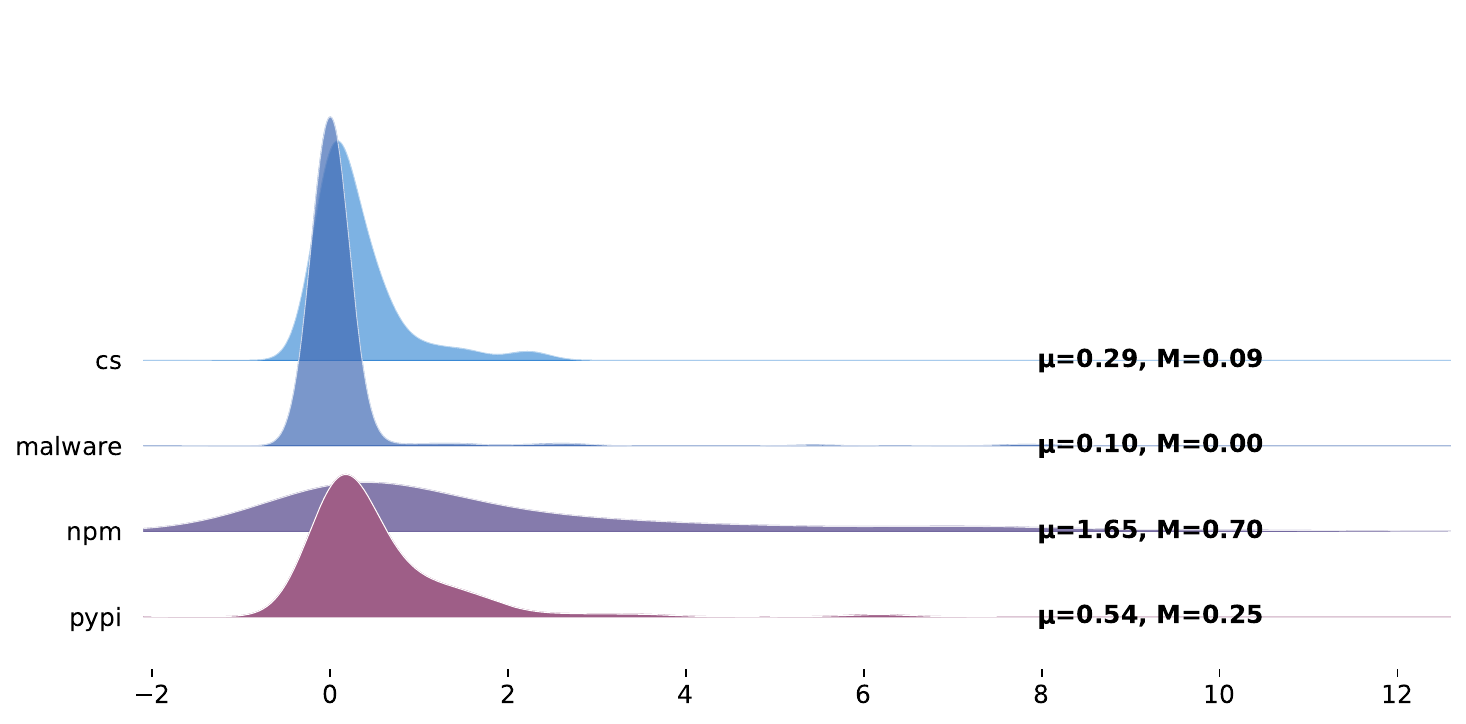}
    \caption{Distribution of closures.}
    \label{fig:distr_nom.closures}
\end{subfigure}

\caption{Distribution of metrics per code category (continued). Notation: $\mu$ denotes mean, and M denotes median.}

\end{figure}

Figure \ref{fig:prod} illustrates that all four code categories; malware samples, \texttt{npm} and \texttt{pypi} packages, and the small set of \texttt{cs}, exhibit very similar labor profiles once effort is normalized. The median average staff size ranges from 0.16 FTE (\texttt{npm}) to 0.32 FTE (\texttt{cs}), confirming that these artifacts are typically produced by a single developer working part-time rather than by organized teams. More importantly, the median productivity stabilizes at around 0.45 KLOC/PM for every category, with a maximum variance of only $0.01$ KLOC/PM.

\begin{figure}[!ht]
\centering
\begin{subfigure}[ht]{0.49\textwidth}
\includegraphics[width=\textwidth]{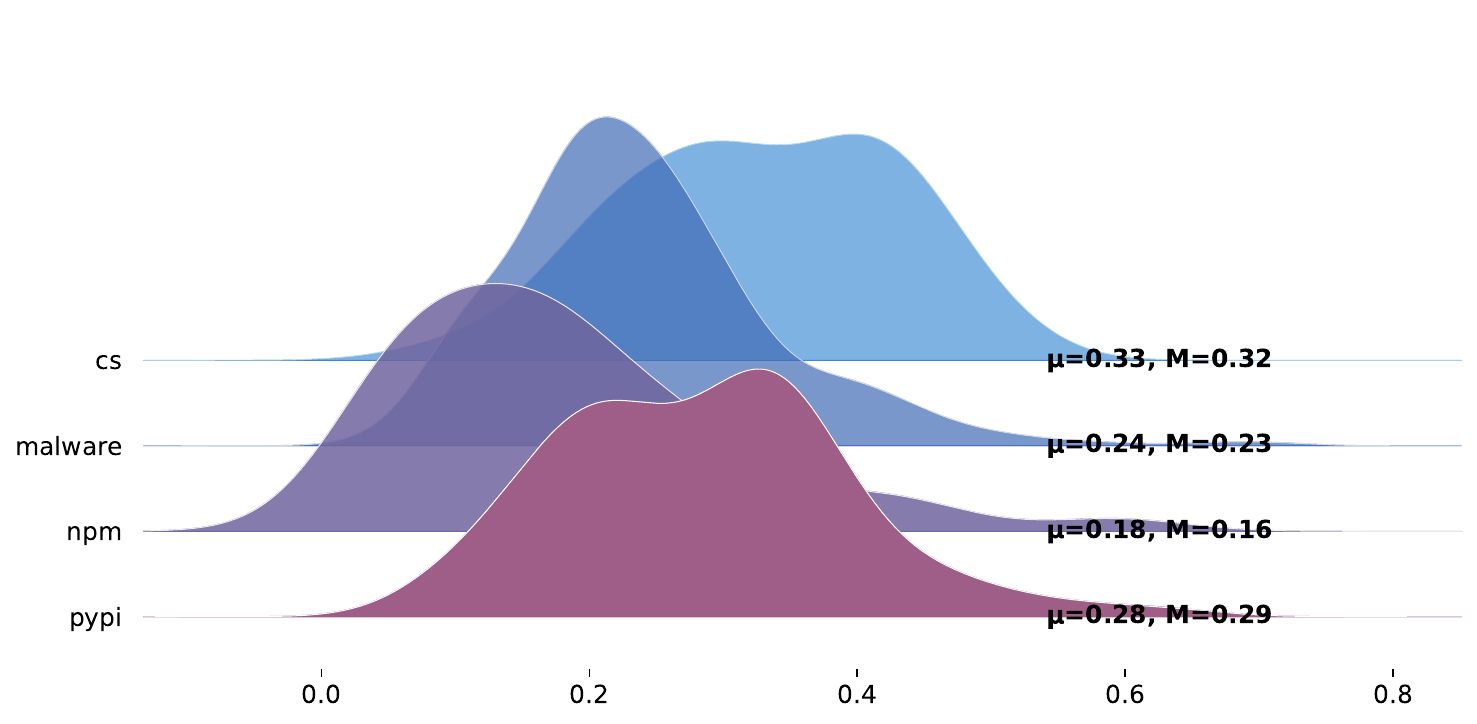}
\caption{Average staff.}
\label{fig:Avg_Staff}
\end{subfigure}\hfill
\begin{subfigure}[ht]{0.49\textwidth}    \centering
\includegraphics[width=\textwidth]{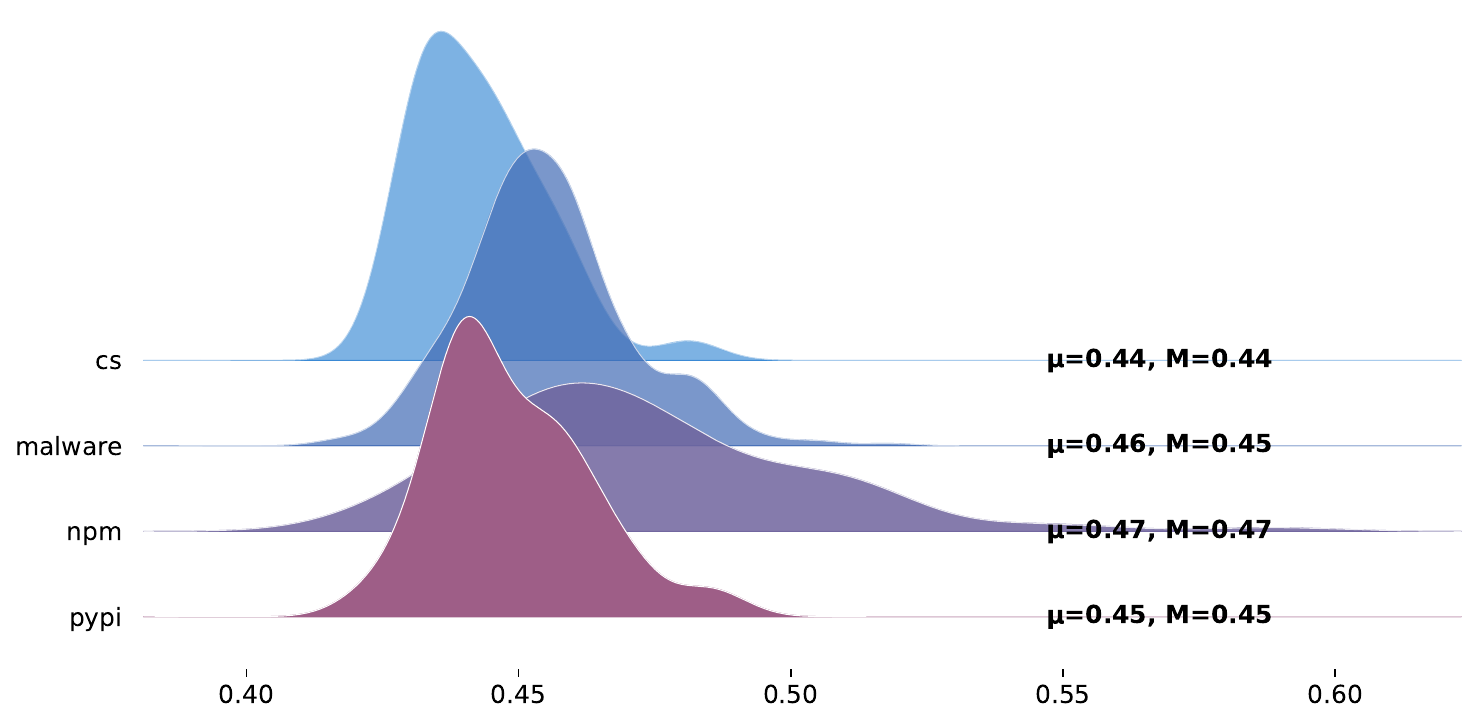}
\caption{Productivity distribution.}
\label{fig:distr_kde_Productivity}
\end{subfigure}
\caption{Distribution of the COCOMO-based productivity and average staff.}
\label{fig:prod}
\end{figure}

To quantify the effort measurement from another perspective, in Figure \ref{fig:Putnam_SLIM1} we show the Putnam-SLIM effort-time trade-off for the four code categories of our dataset.  In our computations, we have used a uniform state-of-technology constant $C_k=2$ as we consider it to be more realistic for the most open source projects of our dataset, as they are not automated and do not follow professional release engineering practices to justify a bigger value. Since $C_k$ is fixed, the only determinant of the vertical offset among the curves is median size. Consequently, the \texttt{npm} sample ($0.13$ KLOC) is lowest, while the \texttt{cs} sample ($0.31$ KLOC) lies highest; the malware and \texttt{pypi} categories fall between. For projects this small, the absolute effort predicted by SLIM is minimal; compressing the schedule to a single month still yields median values well below 0.02 person-months ($\approx$ 3 staff-hours). Then, Figure \ref{fig:Putnam_SLIM2} shows the corresponding average-staff requirement; staffing exceeds 0.05 FTE only when the schedule is driven below ten calendar days. These shapes are consistent with the widely reported `lone-developer' nature of both hobbyist open-source work and the conceptualized model for malware authors.
\begin{figure}[!ht]
\centering
\begin{subfigure}[ht]{0.49\textwidth}
\includegraphics[width=\textwidth]{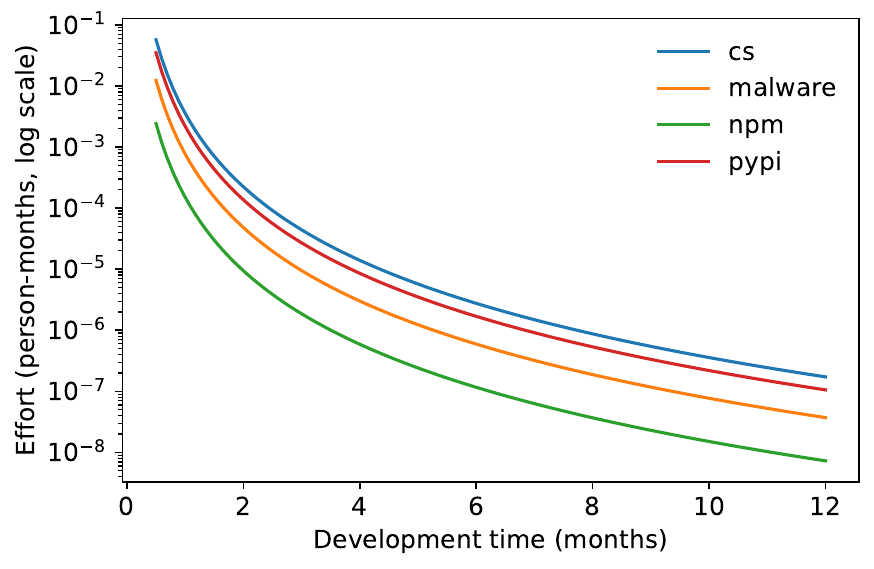}
\caption{Putnam-SLIM effort-time trade-off for the four code categories.}
\label{fig:Putnam_SLIM1}
\end{subfigure}\hfill
\begin{subfigure}[ht]{0.49\textwidth}    \centering
\includegraphics[width=\textwidth]{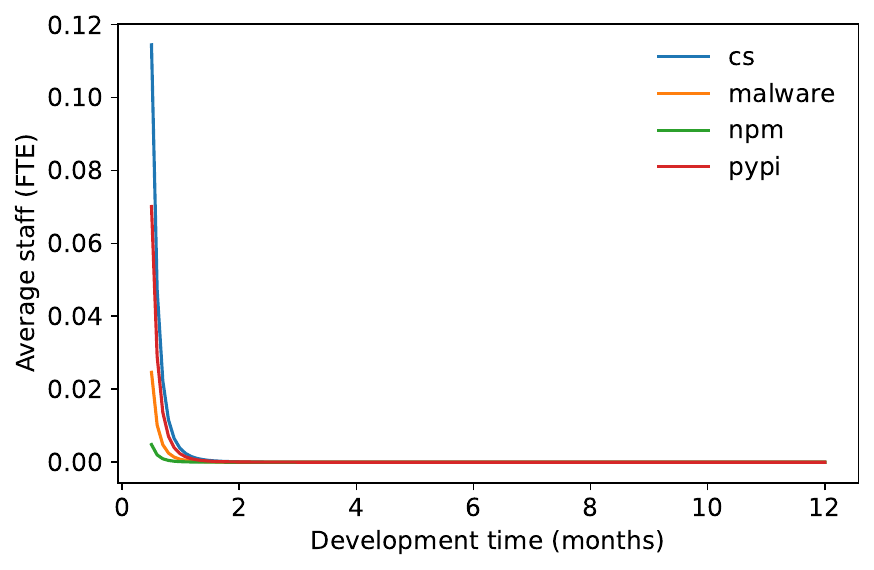}
\caption{Average-staff requirement.}
\label{fig:Putnam_SLIM2}
\end{subfigure}
\caption{Putnam-SLIM model curves for $C_k=2$.}
\label{fig:Putnam_SLIM}
\end{figure}

Finally, we compare the SLIM curves with the COCOMO-based average staff and the observed productivity in KLOC per person-month. Notably, all four categories cluster tightly around 0.16 - 0.32 FTE and exhibit a very similar estimated productivity of approximately 0.45 KLOC/PM. Thus, once size is normalized, malware authors do not appear to be intrinsically faster coders; their perceived speed can be attributed to targeting a significantly smaller functional scope, so we can claim that malware development is not accelerated by superior tooling but by minimizing functionality that would be expected in traditional software.

Finally, to assess the distinguishability of the groups based on the code metrics reported in Table \ref{tbl:metrics}, we analysed the latent structure of the data by applying a Gaussian Mixture Model (GMM), which is well suited to capture complex and potentially overlapping patterns in multidimensional data \cite{patel2020clustering}. To determine the most appropriate number of clusters, we used the Weighted Purity Score to maximise the dominance of the majority class within each cluster \cite{manning2008introduction}. It is important to note that these clustering outcomes serve to highlight structural hints and overlapping tendencies rather than demonstrating strong, definitive separability. GMMs can be sensitive to initialization and parameter configurations, such as covariance type constraints. We then projected all samples in the dataset, including the benign subsets and the malware projects, into a three-dimensional Principal Component Analysis (PCA) space for visualization purposes. The resulting distribution is shown in Figure \ref{fig:clustering}. As can be observed, the clustering reveals the internal structure of the dataset, showing some clear tendencies (e.g., most of malware samples, represented by cross values, belong to the same cluster). While our formulation yields stable dominant classes within the presented clusters, the boundary overlaps emphasize that malicious and benign projects often share fundamental coding patterns, indicating that a perfectly separable structural signature across all software types remains elusive.

\begin{sidewaystable}[!th]
\tiny
\centering
\rowcolors{2}{}{gray!10}
\setlength{\tabcolsep}{4pt} 
\renewcommand{\arraystretch}{0.8}
\begin{tabular}{lrrrr}
\toprule
& \textbf{malware} & \textbf{npm} & \textbf{cs} & \textbf{pypi} \\
\rowcolor{white}\textbf{Metric} &  min, max, avg, std &  min, max, avg, std & min, max, avg, std  & min, max, avg, std  \\
\midrule
metrics.abc.assignments & 0, 27, 0.00, 0.22, 1.98 & 0, 0, 0.00, 0.00, 0.00 & 0, 31, 0.00, 1.50, 6.09 & 0, 17, 0.00, 0.19, 1.77 \\
metrics.abc.branches & 0, 56, 0.00, 0.51, 4.45 & 0, 0, 0.00, 0.00, 0.00 & 0, 78, 0.00, 4.79, 18.23 & 0, 57, 0.00, 0.63, 5.99 \\
metrics.abc.conditions & 0, 34, 0.00, 0.24, 2.25 & 0, 0, 0.00, 0.00, 0.00 & 0, 19, 0.00, 1.01, 4.01 & 0, 14, 0.00, 0.16, 1.55 \\
metrics.cyclomatic.average & 1, 16, 2.58, 2.98, 1.74 & 1, 9, 2.01, 2.46, 1.44 & 1, 8, 2.66, 2.87, 1.33 & 1, 9, 2.40, 2.55, 0.99 \\
metrics.halstead.N1 & 18, 2693, 413.00, 472.50, 338.32 & 3, 1582, 251.04, 344.91, 337.04 & 100, 1404, 719.42, 731.16, 355.08 & 35, 1049, 345.75, 383.70, 244.15 \\
metrics.halstead.N2 & 26, 1629, 324.00, 386.18, 267.29 & 1, 1108, 179.58, 270.81, 266.05 & 56, 1169, 521.28, 566.17, 269.16 & 42, 1258, 462.25, 488.62, 287.47 \\
metrics.halstead.bugs & 0, 5, 1.07, 1.28, 0.95 & 0, 4, 0.64, 1.03, 1.13 & 0, 3, 2.27, 2.14, 1.04 & 0, 3, 1.37, 1.42, 0.94 \\
metrics.halstead.difficulty & 2, 86, 25.34, 27.49, 13.28 & 1, 124, 23.15, 27.17, 19.87 & 8, 65, 34.69, 35.53, 11.79 & 6, 60, 29.55, 29.71, 11.97 \\
metrics.halstead.effort & 463, 3175049, 338455.94, 495292.34, 569168.44 & 29, 2304854, 111301.66, 339513.53, 500591.31 & 14535, 2737251, 1352715.18, 1294142.04, 811856.77 & 6472, 2895214, 441172.75, 615143.39, 639998.37 \\
metrics.halstead.estimated\_program\_length & 71, 9084, 804.13, 970.79, 798.19 & 12, 2301, 529.20, 664.98, 575.67 & 185, 6782, 999.81, 1243.91, 1061.74 & 165, 3328, 858.82, 992.75, 600.56 \\
metrics.halstead.length & 53, 3695, 731.15, 858.68, 581.08 & 7, 2690, 448.26, 615.72, 600.67 & 157, 2519, 1240.70, 1297.33, 620.78 & 77, 2307, 830.84, 872.32, 522.10 \\
metrics.halstead.level & 0, 0, 0.13, 0.14, 0.09 & 0, 0, 0.08, 0.10, 0.08 & 0, 0, 0.10, 0.11, 0.04 & 0, 0, 0.11, 0.13, 0.10 \\
metrics.halstead.n1 & 2, 40, 15.58, 16.60, 5.76 & 3, 44, 19.00, 19.08, 7.07 & 9, 24, 18.03, 17.99, 3.11 & 4, 24, 16.46, 16.02, 3.89 \\
metrics.halstead.n2 & 17, 870, 98.66, 116.17, 82.08 & 1, 281, 68.70, 80.09, 65.92 & 27, 507, 116.14, 132.46, 78.29 & 24, 336, 107.76, 117.42, 60.42 \\
metrics.halstead.purity\_ratio & 0, 5, 1.40, 1.41, 0.44 & 0, 2, 1.51, 1.50, 0.38 & 0, 1, 1.22, 1.26, 0.16 & 0, 2, 1.52, 1.55, 0.31 \\
metrics.halstead.time & 25, 176391, 18803.11, 27516.24, 31620.47 & 1, 128047, 6183.43, 18861.86, 27810.63 & 807, 152069, 75150.84, 71896.78, 45103.15 & 359, 160845, 24509.60, 34174.63, 35555.47 \\
metrics.halstead.vocabulary & 19, 874, 115.16, 132.76, 83.91 & 6, 287, 84.43, 99.17, 70.58 & 37, 531, 133.18, 150.46, 80.14 & 33, 347, 122.92, 133.44, 62.62 \\
metrics.halstead.volume & 225, 33929, 5644.27, 6916.81, 5268.23 & 19, 22129, 3174.81, 4773.30, 5074.87 & 947, 26582, 10282.00, 11420.60, 6273.91 & 457, 21305, 6577.56, 7217.00, 4866.73 \\
metrics.loc.blank & 0, 148, 22.50, 28.77, 21.43 & 0, 80, 7.35, 14.78, 18.29 & 10, 97, 37.89, 42.60, 18.15 & 5, 153, 40.93, 46.63, 29.19 \\
metrics.loc.cloc & 0, 413, 14.75, 29.47, 43.71 & 0, 568, 16.75, 39.59, 77.03 & 6, 197, 57.11, 63.73, 43.22 & 2, 274, 41.51, 48.66, 39.11 \\
metrics.loc.lloc & 0, 601, 71.56, 88.02, 71.59 & 1, 278, 35.93, 61.80, 65.88 & 18, 229, 133.16, 124.62, 57.61 & 10, 308, 95.00, 99.58, 63.53 \\
metrics.loc.ploc & 13, 874, 130.67, 156.54, 105.28 & 1, 468, 73.88, 96.99, 91.97 & 34, 498, 209.81, 236.34, 106.94 & 21, 487, 182.39, 183.65, 104.79 \\
metrics.loc.sloc & 13, 1049, 183.33, 208.51, 137.28 & 1, 808, 106.50, 150.59, 157.52 & 56, 613, 305.84, 334.62, 144.94 & 40, 868, 259.74, 272.61, 154.04 \\
metrics.mi.mi\_original & -19, 134, 55.26, 52.42, 20.13 & -19, 155, 61.84, 62.80, 32.33 & 10, 82, 39.82, 41.65, 14.09 & -6, 120, 43.83, 45.22, 20.49 \\
metrics.mi.mi\_sei & -77, 120, 26.38, 22.58, 27.51 & -55, 148, 40.31, 42.75, 42.82 & -19, 65, 17.62, 19.61, 17.53 & -34, 108, 19.47, 20.77, 26.52 \\
metrics.mi.mi\_visual\_studio & 0, 80, 33.10, 32.20, 10.68 & 0, 90, 36.38, 38.03, 17.64 & 19, 48, 27.27, 28.85, 6.15 & 14, 72, 29.15, 30.21, 9.48 \\
metrics.nargs.average & 0, 4, 0.87, 0.98, 0.64 & 0, 2, 0.74, 0.81, 0.46 & 0, 3, 1.17, 1.38, 0.66 & 0, 3, 1.27, 1.30, 0.60 \\
metrics.nargs.average\_closures & 0, 1, 0.00, 0.01, 0.10 & 0, 2, 0.08, 0.25, 0.42 & 0, 0, 0.02, 0.06, 0.08 & 0, 0, 0.08, 0.12, 0.13 \\
metrics.nargs.average\_functions & 0, 4, 0.87, 0.98, 0.64 & 0, 2, 0.78, 0.85, 0.48 & 0, 3, 1.17, 1.38, 0.66 & 0, 3, 1.28, 1.31, 0.60 \\
metrics.nargs.closures\_max & 0, 1, 0.00, 0.02, 0.14 & 0, 6, 0.14, 0.48, 0.82 & 0, 0, 0.05, 0.13, 0.20 & 0, 2, 0.14, 0.26, 0.40 \\
metrics.nargs.functions\_max & 0, 6, 1.59, 1.75, 1.11 & 0, 5, 1.41, 1.58, 1.06 & 1, 6, 2.33, 2.60, 1.18 & 0, 7, 2.56, 2.76, 1.41 \\
metrics.nargs.total & 0, 67, 4.98, 6.87, 7.59 & 0, 69, 6.00, 10.61, 13.04 & 2, 51, 16.72, 19.65, 11.43 & 0, 74, 24.28, 25.65, 16.31 \\
metrics.nargs.total\_closures & 0, 6, 0.00, 0.07, 0.50 & 0, 11, 0.15, 1.39, 2.48 & 0, 2, 0.08, 0.30, 0.50 & 0, 6, 0.18, 0.54, 0.92 \\
metrics.nargs.total\_functions & 0, 65, 4.88, 6.79, 7.46 & 0, 59, 5.00, 9.22, 11.59 & 1, 51, 16.50, 19.35, 11.30 & 0, 72, 23.95, 25.11, 16.12 \\
metrics.nexits.sum & 0, 51, 4.64, 6.09, 6.43 & 0, 50, 5.00, 9.54, 12.18 & 1, 21, 12.34, 12.12, 5.37 & 0, 51, 6.77, 8.57, 7.17 \\
metrics.nom.average & 0, 1, 0.32, 0.35, 0.19 & 0, 0, 0.46, 0.50, 0.24 & 0, 0, 0.48, 0.49, 0.13 & 0, 1, 0.62, 0.59, 0.18 \\
metrics.nom.closures & 0, 8, 0.00, 0.10, 0.70 & 0, 10, 0.70, 1.65, 2.33 & 0, 2, 0.09, 0.29, 0.48 & 0, 6, 0.25, 0.54, 0.87 \\
metrics.nom.closures\_average & 0, 0, 0.00, 0.01, 0.04 & 0, 0, 0.05, 0.09, 0.11 & 0, 0, 0.00, 0.02, 0.03 & 0, 0, 0.01, 0.03, 0.08 \\
metrics.nom.functions & 0, 33, 3.13, 4.07, 3.90 & 0, 41, 3.58, 6.97, 8.75 & 2, 36, 9.13, 11.13, 7.11 & 0, 47, 12.69, 14.94, 9.94 \\
metrics.nom.functions\_average & 0, 0, 0.32, 0.34, 0.19 & 0, 0, 0.39, 0.41, 0.22 & 0, 0, 0.46, 0.47, 0.13 & 0, 0, 0.59, 0.56, 0.15 \\
metrics.nom.total & 0, 35, 3.15, 4.18, 4.18 & 0, 44, 5.00, 8.62, 10.12 & 2, 36, 9.95, 11.42, 7.27 & 0, 48, 12.93, 15.48, 10.24 \\
metrics.npa.average & 0, 0, 0.00, 0.00, 0.02 & 0, 0, 0.00, 0.00, 0.00 & 0, 0, 0.00, 0.01, 0.02 & 0, 0, 0.00, 0.00, 0.00 \\
metrics.npa.class\_attributes & 0, 27, 0.00, 0.12, 1.50 & 0, 0, 0.00, 0.00, 0.00 & 0, 9, 0.00, 0.45, 1.84 & 0, 2, 0.00, 0.03, 0.27 \\
metrics.npa.classes & 0, 25, 0.00, 0.08, 1.33 & 0, 0, 0.00, 0.00, 0.00 & 0, 2, 0.00, 0.09, 0.40 & 0, 0, 0.00, 0.01, 0.05 \\
metrics.npa.classes\_average & 0, 0, 0.00, 0.00, 0.02 & 0, 0, 0.00, 0.00, 0.00 & 0, 0, 0.00, 0.01, 0.02 & 0, 0, 0.00, 0.00, 0.00 \\
metrics.npa.interface\_attributes & 0, 1, 0.00, 0.00, 0.06 & 0, 0, 0.00, 0.00, 0.00 & 0, 0, 0.00, 0.00, 0.00 & 0, 0, 0.00, 0.00, 0.00 \\
metrics.npa.interfaces & 0, 1, 0.00, 0.00, 0.06 & 0, 0, 0.00, 0.00, 0.00 & 0, 0, 0.00, 0.00, 0.00 & 0, 0, 0.00, 0.00, 0.00 \\
metrics.npa.interfaces\_average & 0, 0, 0.00, 0.00, 0.00 & 0, 0, 0.00, 0.00, 0.00 & 0, 0, 0.00, 0.00, 0.00 & 0, 0, 0.00, 0.00, 0.00 \\
metrics.npa.total & 0, 25, 0.00, 0.09, 1.33 & 0, 0, 0.00, 0.00, 0.00 & 0, 2, 0.00, 0.09, 0.40 & 0, 0, 0.00, 0.01, 0.05 \\
metrics.npa.total\_attributes & 0, 27, 0.00, 0.12, 1.51 & 0, 0, 0.00, 0.00, 0.00 & 0, 9, 0.00, 0.45, 1.84 & 0, 2, 0.00, 0.03, 0.27 \\
metrics.npm.average & 0, 0, 0.00, 0.01, 0.07 & 0, 0, 0.00, 0.00, 0.00 & 0, 0, 0.00, 0.05, 0.17 & 0, 0, 0.00, 0.00, 0.01 \\
metrics.npm.class\_methods & 0, 12, 0.00, 0.10, 0.90 & 0, 0, 0.00, 0.00, 0.00 & 0, 13, 0.00, 0.76, 2.83 & 0, 10, 0.00, 0.11, 1.06 \\
metrics.npm.classes & 0, 9, 0.00, 0.07, 0.68 & 0, 0, 0.00, 0.00, 0.00 & 0, 8, 0.00, 0.52, 1.92 & 0, 8, 0.00, 0.09, 0.86 \\
metrics.npm.classes\_average & 0, 0, 0.00, 0.01, 0.06 & 0, 0, 0.00, 0.00, 0.00 & 0, 0, 0.00, 0.04, 0.16 & 0, 0, 0.00, 0.00, 0.01 \\
metrics.npm.interface\_methods & 0, 0, 0.00, 0.00, 0.03 & 0, 0, 0.00, 0.00, 0.00 & 0, 0, 0.00, 0.03, 0.10 & 0, 1, 0.00, 0.02, 0.18 \\
metrics.npm.interfaces & 0, 0, 0.00, 0.00, 0.03 & 0, 0, 0.00, 0.00, 0.00 & 0, 0, 0.00, 0.03, 0.10 & 0, 1, 0.00, 0.02, 0.18 \\
metrics.npm.interfaces\_average & 0, 0, 0.00, 0.00, 0.01 & 0, 0, 0.00, 0.00, 0.00 & 0, 0, 0.00, 0.01, 0.02 & 0, 0, 0.00, 0.00, 0.00 \\
metrics.npm.total & 0, 9, 0.00, 0.07, 0.69 & 0, 0, 0.00, 0.00, 0.00 & 0, 8, 0.00, 0.54, 2.00 & 0, 10, 0.00, 0.11, 1.04 \\
metrics.npm.total\_methods & 0, 12, 0.00, 0.10, 0.92 & 0, 0, 0.00, 0.00, 0.00 & 0, 13, 0.00, 0.79, 2.92 & 0, 11, 0.00, 0.13, 1.24 \\
metrics.wmc.classes & 0, 42, 0.00, 0.29, 2.77 & 0, 0, 0.00, 0.00, 0.00 & 0, 26, 0.00, 1.56, 6.03 & 0, 22, 0.00, 0.24, 2.30 \\
metrics.wmc.interfaces & 0, 0, 0.00, 0.00, 0.03 & 0, 0, 0.00, 0.00, 0.00 & 0, 0, 0.00, 0.03, 0.10 & 0, 1, 0.00, 0.02, 0.18 \\
metrics.wmc.total & 0, 42, 0.00, 0.29, 2.78 & 0, 0, 0.00, 0.00, 0.00 & 0, 26, 0.00, 1.59, 6.12 & 0, 23, 0.00, 0.26, 2.48 \\
\bottomrule
\end{tabular}

\caption{Computed metrics for each subset of the dataset.}
\label{tbl:metrics}

\end{sidewaystable}

\begin{figure}[!ht]
    \centering
    \includegraphics[width=0.9\textwidth]{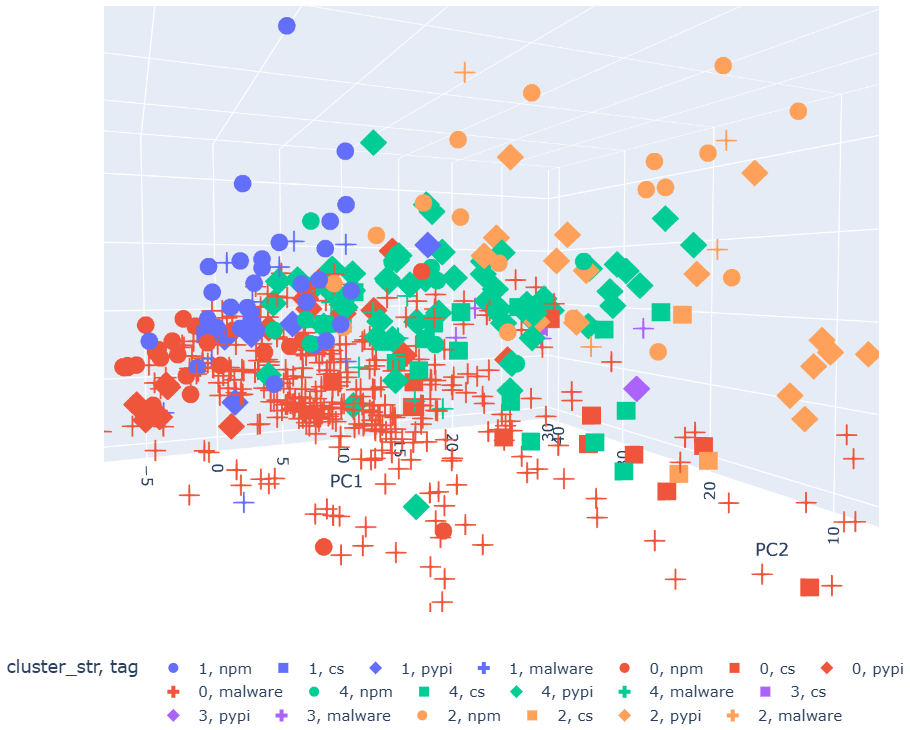}

    \vspace{0.5em}
    \footnotesize
    \begin{tabular}{crr}
        \toprule
        Cluster & Benign & Malware \\
        \midrule
        0 & 55 & 342 \\
        1 & 37 & 11 \\
        2 & 41 & 4 \\
        3 & 3 & 5 \\
        4 & 85 & 7 \\
        \bottomrule
    \end{tabular}
    \caption{Clustering outcomes in a three-dimensional PCA visualization, and cluster composition by simplified tag, showing the dominant class within each cluster.}
    \label{fig:clustering}
\end{figure}

\subsection{Interpreting metrics as behavioral signals}
While programming is often perceived as a purely technical endeavor,  essentially it is a cognitive and behavioral activity. The ways in which developers structure, document, and secure their code reflect their priorities, operational constraints, and underlying development habits and strategies. As discussed, prior research has established that personality traits and cognitive styles shape software development behaviors such as debugging strategies, code reuse, and commenting practices. In this context, malware authorship is particularly revealing. Unlike developers of legitimate software, malware authors operate under different motivational structures, risk tolerances, and ethical constraints, which may influence their development habits.

Our analysis of malware versus benign software reveals several behavioral indicators. For instance, the prevalence of poorly documented and sparsely commented malware implies that malware authors deliberately choose to obfuscate intent and minimize traceability, possibly aligning with a cybersecurity mindset consistent with priorities such as deception, secrecy, and operational efficiency. Moreover, the observed higher cyclomatic complexity and low modularization may indicate a preference for rapid, goal-directed coding rather than adherence to formal software engineering practices. After all, malware development does not follow corporate maintenance workflows, which can reduce incentives for long-term maintainability. Conversely, benign software projects typically exhibit better modularization, higher documentation density, and greater maintainability, reflecting a culture of collaboration, transparency, and long-term usability. Nonetheless, some of the observed patterns may reflect both deliberate choices and situational constraints rather than stable individual traits. For instance, people are expected to react differently when they know that their actions are monitored and assessed by others. As such, they may code differently, because others will see their code, the quality of their work would be assessed on the quality of their code and comments, but also because maintainability is a work requirement.

While these software metrics provide a window into development habits, a critical distinction must be made between a developer's inherent cognitive style and the behavioral adaptations necessitated by a project's lifecycle. The technical signatures identified, such as reduced documentation and increased functional complexity, may not reflect solely a static "malicious mindset," but rather the functional requirements of software designed to be "disposable." Because malware is often optimized for immediate operational success and the minimization of a forensic footprint, its development priorities naturally favor expedience and evasion over the long-term maintainability and transparency required in collaborative benign ecosystems. Therefore, these patterns are best interpreted as manifestations of developer priorities shifting under the specific situational constraints of secrecy, legal risk, and time pressure.

\section{Lessons Learned and Concluding Remarks}
\label{sec:conclusions}

The systematic examination of malware vulnerabilities encourages further research into exploiting those flaws for defense, which can be considered a counterintuitive angle. Typically, researchers examine how malware exploits vulnerabilities in other software; however, we focus on examining vulnerabilities in the malware itself. This approach has been explored previously in narrower settings. For example, the Malvuln project, launched in 2021, pioneered this insight by systematically finding and cataloging security bugs within malware samples. Moreover, reverse engineers strive to find vulnerabilities in how ransomware authors generate their keys to create decryptors \cite{nomore,KimKBKK25} or, in the case of Wannacry, its hard-coded kill switch. Nevertheless, these efforts are typically case-study driven and do not provide a comprehensive picture. We argue that by cataloging common weakness patterns in malware, we lay the groundwork for systematically identifying such `kill switches' or weaknesses in future malware. Thus, if specific vulnerability patterns frequently appear in malware, defenders could possibly disrupt some malware campaigns by exploiting the vulnerability to crash the malware or take control of its C2 server, as in the case of Emotet and Torpig \cite{Stone-GrossCCGSKKV09}.

The fact that SAST has matured as a technique for finding bugs in source code over the past few years can significantly facilitate this process. Traditionally, SAST is used in typical software development to catch bugs early, but it has rarely been applied to malicious code, as malware source code is seldom available to researchers. Notable exceptions exist on a small scale (e.g., analyses of a few botnets or exploit kits in prior work), but until recently, there was no large-scale dataset of malware source code for such a study, as noted in the literature review in Section \ref{sec:related}. In this regard, previous work has tended to measure either code quality in isolation \cite{Calleja19} or the presence of secure-coding flaws independently \cite{Kotov13}. By running an industrial SAST pipeline over the exact same corpus for which we compute size, complexity, and maintainability metrics, we can observe both dimensions simultaneously and quantify their interaction within the same corpus.

While one might expect malware to be sloppily written due to factors such as small teams without a substantial budget, a lack of a traditional client, and often relying on copy-and-paste from other code repositories, our results show the opposite. We observe only modestly higher cyclomatic complexity in malware functions on average compared to benign software, and malware's maintainability index values were often in a similar range to those of benign code. In fact, malware samples are often small and focused, which can help maintain high specific metric scores. Nevertheless, there are notable outliers, e.g., some malware pieces showed extremely high Halstead complexity or very low maintainability due to obfuscation or dense, undocumented code. These extremes likely correspond to heavily obfuscated samples or autogenerated code that have no equivalent in typical open-source projects but do not change the overall picture; some malware has good quality, while some is intentionally convoluted. We argue that since malware has become a commodity in the Malware-as-a-Service model \cite{maas}, its quality, even in terms of writing, is and will be further improved to justify the `investments' that are being made on top of it \cite{Calleja19}.

Software metrics are essential tools for assessing and improving the quality, efficiency, and maintainability of software systems. They encompass a broad range of measures offering unique insights into specific aspects of software development. One of the key benefits of software metrics is their ability to provide quantitative measures of various software attributes, including complexity, maintainability, reliability, and development effort. These metrics are often simple to calculate and can be automated, making them efficient tools for integration into development workflows. Metrics like Halstead and function point analysis are language-independent, allowing their application across diverse programming environments and languages. Moreover, they standardize software evaluation, enabling comparisons across projects or iterations and supporting informed decision-making in project scheduling, resource allocation, and quality assurance. Metrics also play a crucial role in error prediction, maintenance effort estimation, and risk management, making them valuable during both the development and post-deployment phases.

Despite their utility, software metrics are not without limitations. Many rely on complete source code to achieve accurate results, making them less effective during early development stages. Their predictive capabilities are often limited by underlying assumptions and simplifications that may not fully capture the complexity of modern software systems. For instance, metrics like Halstead focus heavily on syntactic elements, potentially overlooking critical factors. Additionally, the applicability of metrics may vary significantly depending on the type of software; for example, different types of malware with different behavioral characteristics may require specialized measures \cite{abusitta2021malware}. Over-reliance on automated tools can also lead to misinterpretation if the metrics are not contextualized within the broader development environment. Moreover, achieving a balance between simplicity and accuracy often requires trade-offs that limit the precision of certain metrics. Overall, while outliers can be clearly identified, and some tendency can be observed, code metrics alone cannot determine whether a piece of code exhibits a distinguishable behavior in general, as benign projects share many characteristics with malware ones, as seen in Figure \ref{fig:clustering}. This overlap underscores a central challenge in behavioral cybersecurity: the same "lone-developer" coding style found in hobbyist open-source projects is often mirror-imaged in malware creation, indicating that coding habits are deeply influenced by shared technical constraints regardless of the project's ultimate purpose.

From a scientific perspective, this analysis provides a novel approach to understanding malware development practices by focusing on code quality and security findings rather than traditional dynamic behavior analysis. The consistency of SAST findings within malware families, coupled with distinct patterns between different families, suggests that static code analysis could serve as an additional dimension for malware classification and family attribution. This observation has significant implications for malware analysis methodology, as it indicates that SAST tools, traditionally used to detect vulnerabilities in benign software, can provide valuable insights into malware development practices and potentially aid malware family classification. Note that more than two-fifths of the CWEs we detect originate in code fragments that appear directly in multiple families, implying that a subtle hardening effort focused on a handful of shared libraries could eliminate a large percentage of flaws.

These findings contribute to our understanding of malware development practices and suggest that malicious code development, despite its nefarious purposes, follows discernible patterns that can be detected through standard software analysis tools. This insight could prove valuable for improving malware detection and classification systems by incorporating static analysis patterns as additional features for identification and categorization. However, while structural metrics (e.g., complexity, documentation density, abstraction usage) are informative for comparative characterization across groups, reliable distinction in ambiguous cases requires additional context beyond code metrics alone, such as operational artifacts, campaign linkage, and external threat-intelligence signals, which unfortunately, is not part of the existing dataset. We acknowledge this as a limitation of the current study and a priority direction for future work.

Beyond the latter, our research highlights that in terms of mentality, programmers generate code with certain structural differences. However, they are not capable of accurately distinguishing between the uses. On the other hand, the prioritization, especially in terms of security, is radically different. We consider this a major finding, as in many instances, the individuals who write malicious software may also write benign code. For instance, the notorious advanced persistent threat (APT) group FIN7 tried to recruit coders and hackers using a fake company profile and lure benign software developers\footnote{\url{https://techcrunch.com/2021/10/21/fin7-fake-company-recruit-hacks/}}.

Although we do not support the concept of `hack-back' as it raises many legal and ethical concerns \cite{halberstam2013hacking,kallberg2015right,holzer2016ethics,pool2017police}, we encourage researchers to consider malware as software artifacts that can be scrutinized and even attacked defensively. Relevant future research lines include addressing current limitations of software metrics, such as enhancing predictive accuracy, broadening applicability to diverse software types, and incorporating measures for emerging attributes like sustainability and ethical considerations, efficiently capturing coding practices and behavior beyond just the benign and malicious paradigm.

A key conceptual distinction in our interpretation is between \emph{development strategy} and \emph{cognitive style}. In this study, development strategy denotes objective-driven engineering decisions shaped by operational constraints, such as rapid deployment, secrecy, low maintainability requirements, and evasion of forensic scrutiny. In contrast, cognitive style refers to broader and comparatively stable individual problem-solving tendencies. Under this distinction, the empirical patterns observed here (e.g., reduced documentation, flatter structure, and limited abstraction in malware projects) are interpreted primarily as strategic adaptations to contextual constraints rather than as direct psychological diagnostics of individual authors. Therefore, our results provide stronger evidence for differences in development strategy across benign and malicious ecosystems, while cognitive-style interpretation remains inferential and should be treated with caution. Paired with the above limitation, the predominant nature of C/C++ in malware in the wild represents another potential issue. Therefore, part of the observed structural differences may reflect language-paradigm effects rather than malicious intent alone. 
Furthermore, differences in language ecosystems remain an important confounder. The prominent use of C and C++ in our malware dataset contrasts with the Python and JavaScript ecosystems of our benign datasets. Establishing a perfect equivalent baseline is challenging because C and C++ do not possess the same concept of centralized, small community projects or package registries (such as \texttt{PyPI} or \texttt{npm}) that can be ranked to provide objective, selective criteria for comparison. Thus, our conclusions regarding structural variations should avoid overgeneralization, as they inevitably encompass some of these inherent ecosystem disparities. Finally, some software metrics were extracted with a single toolchain (\texttt{rust-code-analysis}), which currently does not support C\# and Go. As a result, metric-based analyses were conducted on 463 of 658 malware projects ($>$70\% coverage), and findings of such an experiment should be interpreted as representative of this analyzed subset.

\section*{Author Contribution}
V.V. conceived the project in discussion with C.P. and F.C., designed and performed the experiments and/or simulations, curated the data, carried out the formal analysis, and prepared the figures. V.V. wrote the first draft of the manuscript. C.P. and F.C. provided overall scientific guidance, contributed to the study design and interpretation of the results, secured funding and resources, and supervised the work throughout the project. V.V., C.P., and F.C. contributed to the revision and editing of the manuscript and approved the final version for submission. All authors agree to be accountable for all aspects of the work.

\section*{Funding Declaration}
This work was partially supported by the European Commission under the Horizon Europe Programme, as part of the project SAFEHORIZON (Grant Agreement No. 101168562). This work was partially supported by Ministerio de Ciencia, Innovación y Universidades, Gobierno de España (Agencia Estatal de Investigación, Fondo Europeo de Desarrollo Regional -FEDER-, European Union) under the research grant PID2024-158490OB-C31 ECEEAS. Fran Casino was supported by the Spanish Ministry of Science and Innovation under the ``Ramón y Cajal'' programme (RYC2023-044857-I).

The content of this article does not reflect the official opinion of the European Union. Responsibility for the information and views expressed therein lies entirely with the authors.

The authors would also like to thank VX Underground (\url{vx-underground.org}) for sharing malware code online.

\bibliographystyle{plain}
\bibliography{references}
\end{document}